\documentclass[conference]{IEEEtran}
\IEEEoverridecommandlockouts
\usepackage{subcaption}
\usepackage{subfiles}
\usepackage{import}
\usepackage{enumitem}

\usepackage{tikz}

\usepackage[super]{nth}

\usepackage[raggedrightboxes]{ragged2e}

\usepackage{multirow} 

\usepackage{algorithm}

\usepackage{tabularx}
\usepackage{booktabs}
\usepackage{array}

\usepackage{xurl}

\newcounter{MR}

\usepackage{hyperref}
\hypersetup{
    colorlinks=true,
    linkcolor=blue,
    filecolor=magenta,      
    urlcolor=cyan,
    pdftitle={Overleaf Example},
    pdfpagemode=FullScreen,
    }
    
\usepackage{adjustbox}
\usepackage{csquotes}
\usepackage{wrapfig}
\usepackage{float}

\usepackage{adjustbox}

\usepackage{fontawesome5} 

\usepackage[super]{nth}

\usepackage{lscape}
\usepackage{booktabs}
\usepackage{tabularx}

\usepackage{amsmath,amssymb,amsfonts}
\usepackage{algorithmic}
\usepackage{graphicx}
\usepackage{textcomp}
\usepackage{xcolor}
\def\BibTeX{{\rm B\kern-.05em{\sc i\kern-.025em b}\kern-.08em
    T\kern-.1667em\lower.7ex\hbox{E}\kern-.125emX}}
    
\newcounter{DaveCommentCounter}
\newcounter{MattCommentCounter}
\newcounter{GeorgeAndJiachengCommentCounter}
\newcounter{GabrielCommentCounter}
\usepackage{censor} 
\StopCensoring
    
\begin{document}

\title{Integrating Artificial Intelligence with Human Expertise: An In-depth Analysis of ChatGPT's Capabilities in Generating Metamorphic Relations}

\author{
\IEEEauthorblockN{Yifan Zhang, Dave Towey*\thanks{*Dave Towey is the corresponding author}, Matthew Pike}
    \IEEEauthorblockA{\textit{School of Computer Science}\\
    \textit{University of Nottingham Ningbo China}\\
    Ningbo, Zhejiang 315100\\
    People's Republic of China\\
    \{yifan.zhang, dave.towey, matthew.pike\}@nottingham.edu.cn}
\and
    \IEEEauthorblockN{Quang-Hung Luu, Huai Liu, and Tsong Yueh Chen}
    \IEEEauthorblockA{\textit{Department of Computing Technologies}\\
    \textit{Swinburne University of Technology}\\
    1 John Street, Hawthorn, Melbourne, 3122\\
    Victoria, Australia\\
    \{hluu, hliu, tychen\}@swin.edu.au}
}

\maketitle

\begin{abstract}



\textbf{Context:} This paper provides an in-depth examination of the generation and evaluation of Metamorphic Relations (MRs) using GPT models developed by OpenAI, with a particular focus on the capabilities of GPT-4 in software testing environments.

\textbf{Objective:} The aim is to examine the quality of MRs produced by GPT-3.5 and GPT-4 for a specific System Under Test (SUT) adopted from an earlier study, and to introduce and apply an improved set of evaluation criteria for a diverse range of SUTs.

\textbf{Method:} The initial phase evaluates MRs generated by GPT-3.5 and GPT-4 using criteria from a prior study, followed by an application of an enhanced evaluation framework on MRs created by GPT-4 for a diverse range of nine SUTs, varying from simple programs to complex systems incorporating AI/ML components. A custom-built GPT evaluator, alongside human evaluators, assessed the MRs, enabling a direct comparison between automated and human evaluation methods.

\textbf{Results:} The study finds that GPT-4 outperforms GPT-3.5 in generating accurate and useful MRs. With the advanced evaluation criteria, GPT-4 demonstrates a significant ability to produce high-quality MRs across a wide range of SUTs, including complex systems incorporating AI/ML components.

Conclusions: GPT-4 exhibits advanced capabilities in generating MRs suitable for various applications. The research underscores the growing potential of AI in software testing, particularly in the generation and evaluation of MRs, and points towards the complementarity of human and AI skills in this domain.

\end{abstract}


\section{Introduction}\label{sec:introduction}
\begin{table*}
    \centering
    \caption{Marking table for evaluating the quality of MRs in the earlier study~\cite{STA-chatgpt}}
    \label{tab:STA-marking-scheme}
    \begin{tabularx}{\textwidth}{@{}p{4.5cm}Xc@{}}
        \toprule
        \textbf{Criteria} & \textbf{Description} & \textbf{Score} \\
        \midrule
        Correctness & Does the MR accurately capture the intended behavior? & 0-5 \\
        Applicability & Is the MR applicable to a wide range of inputs? & 0-5 \\
        Novelty & Is the MR a new and original idea? & 0-5 \\
        Clarity & Is the MR easy to understand and unambiguous? & 0-5 \\
        Relevance to safety & Does the MR address a safety-critical aspect of the module? & 0-5 \\
        Overall usefulness & How useful is the MR for testing the module? & 0-5 \\
        Computational feasibility & Is it computationally feasible to apply the MR? & 0-5 \\
        \bottomrule
    \end{tabularx}
\end{table*}

In the domain of software testing, it is critical to ensure the correctness and reliability of complex systems, especially in scenarios where traditional test oracles are ineffective or absent~\cite{metamorphic_review_challenges}. A test oracle serves as a reference for the expected result, against which the real output of the software is assessed for comparison~\cite{effective-oracle-jss1}. The lack of test oracle (i.e., oracle problem) is common, making it difficult to verify the software's correctness~\cite{effective-oracle-jss1}. Within this context, Metamorphic Testing (MT) has emerged as a powerful technique to address this problem, relying on Metamorphic Relations (MRs) to verify software through the relationships among different sets of inputs and their corresponding outputs, diverging from traditional practices that emphasize the correctness of individual results~\cite{MET-apollo}. MT has been effectively adopted to identify faults in a wide variety of complex software systems, including Google Maps, Amazon, and Facebook~\cite{Survey_MT}.

Current techniques for generating MRs are normally specific to certain application domains and rely on human intelligence~\cite{metamorphic_review_challenges}. This task is further complicated by the need for MRs to be both reliable in exposing faults and practical in terms of computational resources~\cite{MR_enhance_use}. Despite these challenges, MRs are valuable in scenarios where traditional test oracles are absent, particularly in the testing of complex systems that integrate AI and ML models. In these contexts, the unpredictability of outputs makes conventional testing methods less useful~\cite{metamorphic_driverless_car}.

The identification of new MRs is one of the main challenges in MT and has thus become a significant research focus~\cite{metamorphic_review_challenges}. Traditional approaches to MR generation are typically constrained to specific application domains and heavily reliant on human expertise~\cite{met-ai-review}\cite{predict-MT}. As the scope of MT's application continues its broad expansion, the advent of Large Language Models (LLMs) like ChatGPT introduces a transformative potential in the MR generation~\cite{STA-chatgpt}. These models, developed through extensive training on varied datasets, demonstrate capabilities in text comprehension and generation similar to human interaction, potentially offering substantial contributions to the automation and refinement of tasks in software testing~\cite{llm-gpt-survey}.

ChatGPT, a LLM developed by OpenAI, has garnered significant attention for its wide-ranging capabilities~\cite{STA-chatgpt}. Empirical studies have documented its proficiency in a variety of tasks, including the resolution of mathematical and logical problems~\cite{gpt-math}, as well as its contributions to bioinformatics research~\cite{gpt-bioinfo}. Within the area of software engineering, empirical evidence suggests that ChatGPT can generate source code, formulate test cases, and aid in the debugging of programs~\cite{gpt-code}. However, critical assessments in recent literature have highlighted limitations and instances of failure in ChatGPT's performance~\cite{llm-gpt-survey}. The model's effectiveness in executing complex software engineering tasks, which often require advanced cognitive capabilities, remains a subject of ongoing evaluation and debate.

GPT-3.5 and GPT-4 belong to the current commercial versions of OpenAI's GPT series, each marking significant advancements in AI language processing~\cite{gpts}. GPT-3.5 laid the groundwork with improved language understanding and generation, as the free version of ChatGPT~\cite{gpt-3.5-into}. GPT-4, a more sophisticated and larger model, further refined these capabilities, leading to a more nuanced and contextually aware conversational experience in the latest iterations of ChatGPT~\cite{openai2023gpt4}. The evolution from GPT-3.5 to GPT-4 represents a leap in the ability to create AI that can interact in human-like ways, with ChatGPT showcasing the practical application of these advancements in fields ranging from customer service to education, offering refined, accurate, and context-sensitive interactions~\cite{openai2023gpt4}.

Evaluation of MRs traditionally relies on criteria developed from empirical research and expert insights~\cite{metamorphic_review_challenges}. However, such criteria often lack the depth and breadth needed to effectively assess MRs in the context of increasingly complex software environments. This scarcity of comprehensive evaluation methodologies highlights the need for ongoing research and development in this area. In our previous study~\cite{STA-chatgpt}, we proposed a set of evaluation criteria to evaluate the quality of MRs generated by GPT-3.5 around seven aspects: correctness, applicability, novelty, clarity, relevance to safety, overall usefulness, and computational feasibility.

This paper delves into the capabilities of GPT-3.5 and GPT-4, specifically focusing on their capabilities in generating and evaluating MRs. We commence by comparing the MRs generated by both models for a specific System Under Test (SUT)—the parking function of an autonomous driving system, which was the SUT in our previous study~\cite{STA-chatgpt}. This comparison is based on the established evaluation criteria in that study, enabling us to benchmark the progression from GPT-3.5 to GPT-4.

In the meantime, we introduced a refined set of evaluation criteria, aiming to provide a more comprehensive, effective, and objective assessment of MRs. These updated criteria were then applied to MRs generated by GPT-4 for a larger group of nine SUTs, encompassing both simple and complex AI/ML-involved systems. This was performed by the implementation of a custom GPT evaluator, which, in conjunction with the human evaluators who have several years of experience in MT and software testing, offered a detailed comparison of AI and human capabilities in MR evaluation.

The findings from this study not only demonstrate the advanced capabilities of ChatGPT, especially GPT-4, in software testing and MR generation across a wide array of applications, but also highlight the evolving role of AI in software testing. They underscore the increasing proficiency of AI in generating and evaluating MRs while simultaneously recognizing the irreplaceable value of human expertise in conducting critical and detail-oriented evaluation processes. This cooperation between AI and human expertise is significant in simplifying the generation of MRs and advancing the methodology of MT.

The rest of the paper is organized as follows: Section~\ref{sec:GPT-version-compare} compares the qualities of MRs generated by GPT-3.5 and GPT-4. Section~\ref{sec:evaluation-criteria} outlines the updated evaluation criteria that are more comprehensive and objective. Then Section~\ref{sec:methodology} explains the experiments about asking GPT-4 to generate MRs for new sets of SUTs and configuring a new GPT evaluator to evaluate the MRs along with the human experts, using the same evaluation criteria proposed in the previous section. Section~\ref{sec:math-results} to Section~\ref{sec:AI-results} lists the experiment results, where Section~\ref{sec:discussion-newMR} summarizes the findings. Section~\ref{sec:discussion} discusses the limitations of the study and the advantages and future work to use GPT as MR evaluators. Finally, Section~\ref{sec:conclusion} concludes the paper.

\section{Quality of MRs: GPT-3.5 VS GPT-4}\label{sec:GPT-version-compare}


In this section, we evaluate the MRs generated by the free ChatGPT version (GPT-3.5) and the subscription-based version (GPT-4). The GPT-3.5 model is designed to provide responses based on a wide range of data sources, making it potentially suitable for general inquiries and everyday use~\cite{gpt-3.5-into}. The GPT-4 model, however, excels in delivering more accurate, detailed, and nuanced answers, which is fit for professional and specialised applications where higher quality and precision in AI responses are essential~\cite{openai2023gpt4}.

In our earlier study~\cite{STA-chatgpt}, we examined the quality of MRs from GPT-3.5, specifically focusing on autonomous driving functions. The results demonstrated that ChatGPT can be a cost-effective
and time-saving approach for generating diverse
and relevant MRs. Furthermore, asking ChatGPT to generate a small number of MRs could give better results than generating a large number of MRs~\cite{STA-chatgpt}.

To compare the quality of MRs generated by the GPT-3.5 and GPT-4 models, we gave the same prompts to GPT-4, and evaluated the results using the same evaluation criteria proposed in our earlier study~\cite{STA-chatgpt}, as Table~\ref{tab:STA-marking-scheme} shows. The table includes several criteria, each assessing a different aspect of the MR's quality~\cite{STA-chatgpt}. ``Correctness'' measures how accurately the MR captures the intended behavior of the software or system. ``Applicability'' evaluates the MR's versatility and its suitability for a range of inputs. ``Novelty'' focuses on the uniqueness and originality of the MR, highlighting new and innovative approaches in the field. ``Clarity'' assesses the ease of understanding and implementing the MR, ensuring it is straightforward and unambiguous. ``Relevance to safety'' is concerned with how well the MR addresses critical safety aspects of the module or system under test. ``Overall usefulness'' provides a general evaluation of the MR's benefits in the testing process, including its effectiveness in uncovering errors. Lastly, ``computational feasibility'' examines the practicality of applying the MR in terms of computational resources, favoring those that are efficient and not overly resource-intensive. Each criterion is scored from 0 to 5, allowing a detailed assessment of the MR's strengths and weaknesses in these areas.

\subsection{Generation of MRs for new target systems}
Our previous research indicated that the GPT models yield more effective results when tasked with creating a limited set of MRs~\cite{STA-chatgpt}. To ensure consistency and control validity threats in this study, we followed a similar approach with GPT-4. To gain fair and useful answers from ChatGPT, we used zero-short prompting. Zero-shot prompting involves interacting with the model without providing explicit examples or prior context~\cite{zero-shot}. The operation was performed in two steps: first clarifying with ChatGPT about the target system, and then asking it to generate a set of unique MR candidates for each target system in each independent session. Our previous experiment involved using GPT-3.5 to generate five specific MRs~\cite{STA-chatgpt}. In interaction with GPT-4, by limiting the scope to the generation of five MRs for the same function, we established a controlled environment, allowing for a direct and fair comparison between the outputs of GPT-3.5 and GPT-4. This approach ensures that the changes observed can be attributed more confidently to the GPT versions rather than external variables. The results were evaluated by experts who have several years of experience in both MT and domain knowledge associated with the systems. These are the same experts who previously assessed the MRs generated by GPT-3.5~\cite{STA-chatgpt}.


\subsection{Evaluation results}\label{subsec:gpts-compare-results}

\begin{figure*}
\captionsetup{width=0.95\textwidth}
    \centering
    \includegraphics[width=0.95\textwidth]{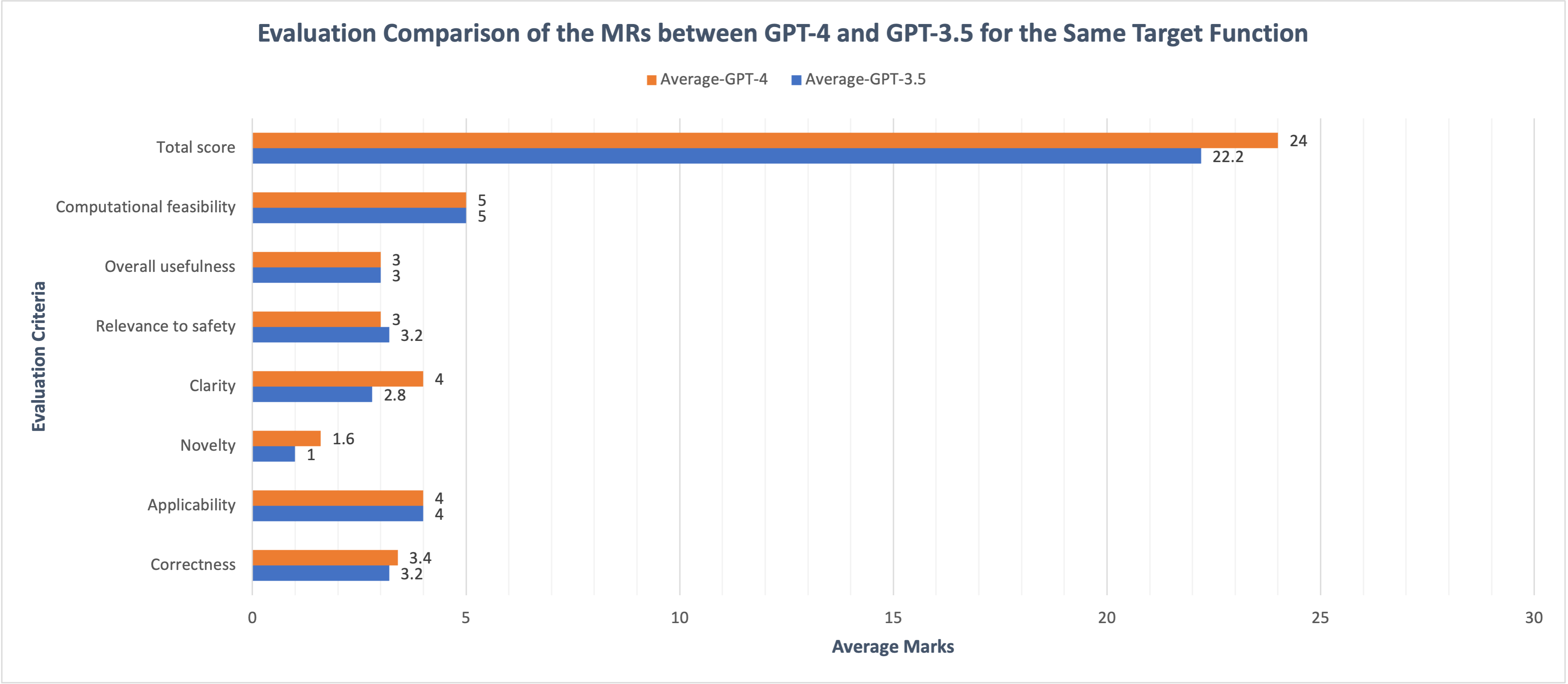}
    \caption{Evaluation comparison chart of MRs for the same target function between GPT-4 and GPT-3.5}
    \label{fig:gpt-evaluation-results-compare}
\end{figure*}

Table~\ref{tab:autonomous-parking-mr} presents the MRs formulated by GPT-4 for the ADS Parking function. These MRs explore different aspects such as changes in parking spot order, weather conditions, obstacle proximity, vehicle dimensions, and time of day. 

\begin{table*}
\centering
\caption{MRs for ADS's parking function generated by GPT-4}
\begin{tabular}{|p{2.3cm}|p{15cm}|}
\hline
\textbf{MRs} & \textbf{Description} \\
\hline
MR1: Permutation of Parking Spot Order & In the original input, the vehicle is instructed to park in a designated spot among a sequence of available spots. The modified input involves permuting the sequence of available parking spots. The expected output is that the vehicle should still successfully park in a designated spot, with the choice of the spot consistent with the vehicle's parking algorithm, irrespective of the permutation of spots. \\
\hline
MR2: Varying Weather Conditions & The original input involves parking under clear weather conditions. The modified input replicates the parking maneuver under different weather conditions (e.g., rain, snow). The expected output is that the vehicle should successfully complete the parking maneuver, adjusting its approach for safety, and ultimately achieving the same parked position. \\
\hline
MR3: Change in Obstacle Proximity & Originally, parking occurs at a certain distance from obstacles. The modified input adjusts the distance from these obstacles. The expected output is that the vehicle adjusts its parking maneuver but still parks successfully and safely in the designated spot. \\
\hline
MR4: Varying Vehicle Dimensions & The original input executes a parking maneuver with a vehicle of specific dimensions. The modified input uses a vehicle of different dimensions. The expected output is that the autonomous system adapts its parking strategy to accommodate the different vehicle size, resulting in a successful and safe park. \\
\hline
MR5: Time of Day Variation & The original input involves performing a parking maneuver during daytime. The modified input performs the same maneuver at night or during twilight. The expected output is that the vehicle uses its sensors and algorithms to successfully and safely complete the parking maneuver, as it did during the daytime. \\
\hline
\end{tabular}
\label{tab:autonomous-parking-mr}
\end{table*}

\begin{table*}
  \centering
  \small 
  \caption{Evaluation results of the MRs generated by GPT-4}
    \begin{tabular}{|p{3.4cm}|p{2.4cm}|p{2.4cm}|p{2.4cm}|p{2.4cm}|p{2.4 cm}|}
    \hline
    \textbf{Marking Criteria} & \textbf{MR1} & \textbf{MR2} & \textbf{MR3} & \textbf{MR4} & \textbf{MR5} \\
    \hline
    Correctness & 3 & 3 & 3 & 4 & 4 \\
    \hline
    Applicability & 4 & 4 & 4 & 4 & 4 \\
    \hline
    Novelty & 3 & 1 & 2 & 1 & 1 \\
    \hline
    Clarity & 4 & 4 & 4 & 4 & 4 \\
    \hline
    Relevance to safety & 2 & 2 & 4 & 3 & 3 \\
    \hline
    Overall usefulness & 3 & 3 & 3 & 3 & 3 \\
    \hline
    Computational feasibility & 5 & 5 & 5 & 5 & 5 \\
    \hline
    Total score & 24 & 22 & 25 & 24 & 24 \\
    \hline
    Explanation & MR1 scores well in applicability and computational feasibility, indicating its practical utility in a variety of testing scenarios. Its correctness and clarity are adequate, aligning it reasonably well with the system's functionalities. However, its novelty is limited, as there are similar MRs in the literature~\cite{permutate-order-mr}. This MR is somewhat lacking in terms of offering innovative approaches or significantly enhancing the safety aspects of the autonomous parking system testing. & MR2 demonstrates high computational feasibility, ensuring its ease of implementation in testing environments. It achieves moderate scores across correctness, applicability, clarity, and relevance to safety, indicating a balanced yet not exceptional contribution to testing methodologies. Its novelty is low, as it is similar to MRs already published in the literature~\cite{change-weather-mr}. This MR could benefit from including a more unique approach of constructing follow-up test scenarios to enhance its usefulness in the testing landscape. & MR3 exhibits high scores in correctness, applicability, clarity, and safety relevance, marking it as a robust and clear MR for testing the parking function. Its computational feasibility is also noteworthy. However, its novelty is also less pronounced. While MR3 is practical and aligns well with safety considerations, its contribution to introducing new testing methodologies is moderate. & MR4 is marked by its high applicability and relevance to safety, underscoring its importance in diverse testing scenarios. It maintains moderate scores in correctness, clarity, and overall usefulness. The novelty of this MR is somewhat limited since it is close to the requirements of the function. While it is effective for testing, its potential for introducing new insights or methods is relatively modest. & MR5 achieves high marks in correctness, applicability, safety relevance, and computational feasibility, with its precision and practicality in safety-critical scenarios. Its novelty, however, is moderate, indicating that the MR does not significantly deviate from the approaches documented in current literature~\cite{change-weather-mr}. The MR's clarity and overall usefulness are satisfactory, but there is room for improvement to maximize its impact and uniqueness in the field of autonomous parking system testing.
  \\
    \hline
    \end{tabular}%
  \label{tab:MR-evaluation-GPT4}%
\end{table*}%

\begin{table*}
\centering
\caption{Evaluation comparison table of MRs for the same target function between GPT-4 and GPT-3.5}
\begin{tabularx}{\textwidth}{@{}p{1.1cm}p{1.3cm}p{1.3cm}p{0.8cm}p{0.8cm}p{2.4cm}p{2.4cm}p{3.2cm}p{2cm}@{}}
\toprule
\textbf{MR} & \textbf{Correctness} & \textbf{Applicability} & \textbf{Novelty} & \textbf{Clarity} & \textbf{Relevance to safety} & \textbf{Overall Usefulness} & \textbf{Computational Feasibility} & \textbf{Total Score} \\
\midrule
\textbf{GPT-3.5} & 3.2 & 4.0 & 1.0 & 2.8 & 3.2 & 3.0 & 5.0 & 22.2 \\
\textbf{GPT-4}   & 3.4 & 4.0 & 1.6 & 4.0 & 3.0 & 3.0 & 5.0 & 24.0 \\

\bottomrule
\end{tabularx}
\label{tab:gpt-evaluation-result-compare}
\end{table*}

The evaluation results are presented in Table~\ref{tab:MR-evaluation-GPT4}. In terms of correctness, the MRs generally display a high level of accuracy and alignment with expected results. ``Applicability'' is a strong area for all MRs, indicating their versatile use in different testing scenarios. However, in terms of ``novelty'', all the MRs generated showed markedly low scores. This suggests that the MRs show many similarities with the ones already established in the literature, indicating a lack of innovative content or original approaches. ``Clarity'' is consistently high across all MRs, meaning that their applications and processes are easily understandable. The MRs' ``relevance to safety'' shows moderate scores, indicating their significance in safety considerations, yet suggesting room for more focused development in this area. Similarly, in terms of ``overall usefulness'', the MRs achieve moderate scores. This indicates their positive impact on testing processes, but implies potential for further enhancement. Lastly, ``computational feasibility'' is a standout feature for all MRs, reflecting their ease of integration and practicality in current testing environments.

Table~\ref{tab:gpt-evaluation-result-compare} includes the average values of scores for MRs. Figure~\ref{fig:gpt-evaluation-results-compare} presents an evaluation comparison of MRs between GPT-4 and GPT-3.5 generated for the same target system. Compared to MRs from GPT-3.5~\cite{STA-chatgpt}, GPT-4 shows a slight improvement in ``correctness''. Both versions score equally in ``applicability'', showing their similar capability to be applied across different scenarios. GPT-4 scores a higher ``novelty'' mark than GPT-3.5, suggesting slightly more innovative approaches compared to GPT-3.5, though the ``novelty'' score itself is not high.

However, ``clarity'' is significantly better in GPT-4, since the MRs generated by GPT-4 have a clear structure and less vague expressions that make the MRs clearer and more comprehensible. This is a significant improvement in GPT-4, as one of the main limitations found in our previous study of GPT-3.5 was that the MRs generated were closer to the system specifications instead of the relations among multiple test cases~\cite{STA-chatgpt}. GPT-4 has clearly specified the source test case, with both input and output relations, which eases the process of generating test cases and analyzing results against MRs.

In ``relation to safety'', GPT-3.5 scores slightly higher, since its MRs were more aligned with safety considerations. Both versions are equal in terms of ``overall usefulness'', suggesting a balanced contribution to the testing process. In ``computational feasibility'', both GPT-4 and GPT-3.5 achieve the full score, reflecting ease of implementation and practicality in testing environments.

The total score reveals that GPT-4, with a score of 24.0, outperforms GPT-3.5, which had a score of 22.2. This comparison indicates that the advancements in GPT-4 have led to improvements in areas like correctness, novelty, and clarity while maintaining high standards in other aspects such as applicability and computational feasibility.

\section{New Objective Evaluation Criteria for MRs Generated by GPT}\label{sec:evaluation-criteria}
\begin{table*}[h]
\centering
\caption{Updated evaluation criteria for MRs}
\label{tab:mr_evaluation_criteria}
\begin{tabularx}{\textwidth}{@{}p{4.5cm}p{10cm}p{2.5cm}@{}}
\toprule
\textbf{Criteria} & \textbf{Criteria Details} & \textbf{Maximum Points} \\ \midrule
Completeness & Includes Source Test Case, Input Relation, Output Relation & 1 \\
Correctness & Assesses Input and Output Relation Adherence, Overall Behavioral Correctness & 3 \\
Generalizability & Evaluates Suitability for Different SUTs: Current, Specific Group, All Types & 3 \\
Novelty & Assesses Originality in Input/Output Relations, Overall Metamorphic Relation & 3 \\
Clarity & Evaluates Clarity Levels: For Experts, Basic Understanding, General Audience & 3 \\
Computational Feasibility & Examines Generation Ease, Automation of Test Cases, MR Validation & 3 \\
Applicability & Focuses on Relation to SUT's Key Features: None, Partial, Strong & 3 \\ \bottomrule
\end{tabularx}
\end{table*}

Our earlier study identified the refinement of evaluation criteria as a potential area for future research. As indicated in Table~\ref{tab:STA-marking-scheme}, the original descriptions for each criterion exhibit a degree of subjectivity and ambiguity. This potentially leads to varying interpretations and assessments by evaluators, each bringing their own unique experiences and knowledge to the evaluation process. Moreover, the criterion ``relevance to safety'' was found to be somewhat restrictive, primarily applicable to SUTs that are related to safety areas, thereby limiting its broader applicability.

In response to these limitations, we have worked on enhancing both the objectivity and the general applicability of these evaluation criteria. The revised set now includes seven criteria: completeness, correctness, generalizability, novelty, clarity, computational feasibility, and applicability of the SUT. Each criterion has been carefully redefined to minimize ambiguity and enhance objectivity, aiming for a more standardized and uniform evaluation process.

In our updated evaluation framework for MRs, each main criterion is divided into several specific sub-criteria, with each contributing to a holistic assessment. This approach not only simplifies the evaluation process but also allows for a more precise and targeted analysis of each component's contribution to the overall functionality and reliability of the MR. The criteria are designed with a scoring system: if an MR completely satisfies all aspects of a criterion, it is awarded full marks. Otherwise, it gains marks for the matching sub-criteria. This approach ensures a clear-cut and straightforward evaluation.

Table~\ref{tab:mr_evaluation_criteria} outlines the evaluation criteria. The ``completeness'' criterion is focused on verifying the inclusion of all essential components within an MR: the source test case, input relation, and output relation. The input relation is necessary for the generation of follow-up test cases, serving as the blueprint for how these cases are derived from the original source test case. Conversely, the output relation contributes to the validation phase. It provides a benchmark against which the outputs of both the source and the follow-up test cases are evaluated, ensuring alignment with the predefined output relation. The ``correctness'' criterion is concerned with how well the input and output relations of the MR adhere to the intended behavioral patterns and the overall correctness of the MR. The updated version improves the evaluation of the completeness and accuracy of MRs by dividing the assessment into separate evaluations of the essential, individual components, which makes the process more detailed and transparent.

The new ``generalizability'' criterion now explicitly defines the range of SUTs for which an MR is suitable (originally the ``applicability'' in the previous criteria, i.e., Table~\ref{tab:STA-marking-scheme}). Unlike the earlier version, which was vague about the ``wide range of inputs'', the new definition categorizes generalizability into three distinct levels: the current SUT; specific groups of SUTs; and all types of SUTs. This structured approach provides a clearer framework for assessing the MR's adaptability, enhancing the precision of our evaluations.

The updated ``novelty'' criterion focuses on assessing the uniqueness of MRs, particularly in their approach and formulation of input/output relations. While it targets the same fundamental aspect as the older version, the refined criterion simplifies the scoring process and enhances objectivity by distinctly evaluating the innovation in input and output relations.

Our new ``clarity'' criterion categorizes the audience into domain experts, individuals with basic understanding, and the general public. This stratification allows evaluators to more objectively determine the MR's comprehensibility across different knowledge levels.

For ``computational feasibility'', the new version introduces three specific dimensions: ease of generating source test cases; feasibility of automating test case generation; and feasibility of automating MR validation. This comprehensive approach, encompassing the entire MT process, provides a more detailed and objective framework compared to the original criterion, which lacked clear definitions for evaluating computational feasibility.

Finally, we have replaced the ``relevance to safety'' criterion with the new ``applicability'' criterion, expanding its applicability beyond safety-focused SUTs. This criterion now evaluates the MR's relevance in three categories: no relevance; partial relevance; and strong relevance to the SUT's key features. Additionally, we have removed the ``Overall Usefulness'' criterion due to its lack of specificity, which made it challenging to apply effectively in assessing the MRs. This revision ensures that each criterion directly contributes to a more targeted and meaningful evaluation of MRs.

The detailed explanations for the new criteria are listed below.

\subsection{Completeness}
The ``completeness'' criterion determines whether the MR includes all essential components.
\begin{itemize}
    \item Presence of key components: The MR must contain the following critical elements (1 mark):
    \begin{itemize}
        \item Source Test Case: Well-defined source test cases that serve as the basis for the MR.
        \item Input Relation: A clear and explicit input relation that specifies how follow-up test cases are generated or modified from the source test case.
        \item Output Relation: A detailed output relation that describes the expected relationships between the outputs of follow-up test cases and the source test cases.
    \end{itemize}

    \item Zero marks for missing components:
    \begin{itemize}
        \item If the MR lacks any of these key components (source test cases, input relation, or output relation), it will be assigned zero marks for completeness.

        \item This criterion is binary: the MR either contains all required components (1 mark) or is missing one or more components (0 marks).

        \item This criterion is considered to be the baseline of MRs. If the MR scores zero marks for completeness, it is considered incomplete or insufficient for evaluation. As a result, it will also receive zero marks under other evaluation criteria.
    \end{itemize}

\end{itemize}

\subsection{Correctness}
The ``correctness'' criterion assesses whether or not the MR accurately represents the intended behavior of the SUT.
\begin{itemize}
    \item If the input relation and output relation violate the SUT’s specifications, this criterion will score zero marks.

    \item The input relation correctness (1 mark):
    \begin{itemize}
        \item The MR's input relation must focus on the current SUT. This means it should not alter the fundamental nature or category of the SUT through its input modifications.
        \item The input relation should explicitly describe how subsequent test cases are derived from the original source test case. This includes clear methodologies or rules for test case generation, ensuring predictability and repeatability.

    \end{itemize}

    \item The output relation correctness (2 marks): 
    \begin{itemize}
        \item The MR's output relation should be free from ambiguous or uncertain language such as `may', `could', or similar terms. It should convey clear and definite relationships.
        \item The output relation must clearly articulate the expected relationships between the outputs of follow-up test cases and the source test cases. This involves specifying the expected changes or consistencies in outputs, based on the input variations.

    \end{itemize}

    \item Overall correctness (3 marks):
    \begin{itemize}
        \item The MR should distinctly focus on the relationships among multiple test cases, rather than mirroring the SUT specifications or assessing the correctness of individual test cases.
        \item This criterion emphasizes the MR's role in revealing systemic behaviors or properties rather than simply reiterating what is already specified about the SUT.

    \end{itemize}

    \item This criterion is also considered to be the baseline of MRs. If the MR scores zero marks for correctness, it is considered incorrect for evaluation. As a result, it will also receive zero marks under other evaluation criteria.
\end{itemize}

\subsection{Generalizability }
This criterion evaluates the MR's suitability across different SUTs.
\begin{itemize}
    
    \item Generalizability to certain SUT (1 mark):
    \begin{itemize}
        \item The MR is specifically tailored to the current SUT and cannot be directly applied to other systems without significant modifications.
        \item The characteristics or requirements of the MR are unique to the current SUT, such as relying on specific hardware, software, or configurations not commonly found in other systems.

    \end{itemize}

    \item Generalizability to certain groups of SUT (2 marks): 
    \begin{itemize}
        \item The MR is generalizable to a specific group of SUTs within a certain type or category. This implies that the MR can be used with systems that share certain characteristics or functionalities but is not universally applicable across all systems of that type.
        \item There exist other SUTs within the same category to which the MR cannot be applied, indicating a level of specificity.

    \end{itemize}

    \item Generalizability to all types of SUT (3 marks):
    \begin{itemize}
        \item The MR can be applied universally to all known SUTs within the same category, demonstrating a high level of adaptability and generalizability.
        \item This level of generalizability implies that the MR is not constrained by specific features, configurations, or special requirements that are unique to certain SUTs.

    \end{itemize}
\end{itemize}

\subsection{Novelty}
The ``novelty'' criterion measures the originality and uniqueness of the MR by comparing it with previously established works and knowledge in the field.
\begin{itemize}
    \item Novelty in the input relation (1 mark):
    \begin{itemize}
        \item The input relation of the MR does not align with any established metamorphic relation input patterns (MRIP)\footnote{MRIP is an abstract representation of the connections between the initial and subsequent inputs within a potentially infinite collection of MRs.~\cite{MR_enhance_use}.}. This means it introduces a new way of altering or creating inputs that are not already documented.
        \item The input relation is distinct and not easily comparable to examples found in existing literature or knowledge in the field. 
    \end{itemize}

    \item Novelty in the output relation (2 marks):
    \begin{itemize}
        \item The output relation of the MR deviates from known metamorphic relation output patterns (MROP)\footnote{MROPs define a set of abstract relationships that focus on the outputs of MRs~\cite{MROP}.}. It suggests an innovative approach to interpreting or utilizing the outputs of an SUT.

    \end{itemize}

    \item Novelty in the overall MR (3 marks):
    \begin{itemize}
        \item The MR as a whole (including both its input and output relations) does not fit within any recognized metamorphic relation patterns (MRPs)\footnote{The MRP describes a set of MRs that share the same level of abstractions~\cite{MR_enhance_use}}. This indicates a completely new approach or method that has not been explored before.
        \item The MR exhibits distinct originality, which indicates no substantial parallels in terms of the selection of source test cases, approach to construct follow-up test cases, and comparison of outputs when compared to existing works and knowledge.
    \end{itemize}

\end{itemize}

\subsection{Clarity}
This criterion assesses how easily the MR can be understood by different audiences.
\begin{itemize}
    \item Clarity for domain experts (1 mark):
    \begin{itemize}
        \item The MR assumes a deep understanding of advanced concepts specific to the field.

        \item It requires familiarity with specialized jargon and technical terminology commonly used among experts

        \item The MR relies on knowledge of specific methodologies and practices typically known only to those with extensive experience or specialized education in the domain.
    \end{itemize}

    \item Clarity for individuals with basic understanding (2 marks): 
    \begin{itemize}
        \item The MR is understandable by individuals with basic knowledge of the field, such as students who have completed relevant introductory courses.

        \item It is also suited for professionals in related fields who possess general knowledge of the topic, but not in-depth specialization.

        \item While extensive experience is not required, the MR assumes the reader has some foundational understanding of the subject.
    \end{itemize}

    \item Clarity for general audience (3 marks): 
    \begin{itemize}
        \item The MR is written to be accessible to a general audience, requiring no specific background knowledge in the field.

        \item Technical terms, if used, are clearly explained to ensure understanding by those unfamiliar with the field.

        \item The MR is structured in a clear, logical manner, making it easy to follow for someone with no prior exposure to the subject matter.
    \end{itemize}

\end{itemize}

\subsection{Computational Feasibility}
The ``computational feasibility'' criterion assesses the ease and practicality of applying the MR in computational terms.
\begin{itemize}
    \item Ease of generating source test cases (1 mark): The MR facilitates the generation of source test cases with minimal complexity. This implies that:
    \begin{itemize}
        \item The process for generating test cases is straightforward and does not require advanced coding skills or specialized software.
        \item It can be accomplished using standard tools and techniques commonly used in the field.
        \item The generation process is efficient, with a low requirement for computational resources and time.
    \end{itemize}

    \item Feasibility of automating test case generation (2 marks): The generation of test cases can be fully automated through code. This level of automation means:
    \begin{itemize}
        \item The process can be executed consistently by a script or program without human intervention.
        \item The automation is replicable, capable of producing consistent results across multiple executions.
        \item The system can adapt to various inputs or changes in requirements with minimal need for reprogramming.

    \end{itemize}

    \item Feasibility of automating MR validation (3 marks): The validation of the MR can be completely automated through code. This involves:
    \begin{itemize}
        \item Developing scripts or programs that can automatically perform the entire process of MT (generating test cases, and comparing outputs against MRs).
        \item The automation is robust, handling different scenarios and edge cases reliably.
        \item The code is efficient in terms of computational resources and time, and operates without the need for manual oversight or intervention for reliable outcomes.
    \end{itemize}

\end{itemize}

\subsection{Applicability}
This criterion evaluates how closely the MR focuses on the key features of the SUT.
\begin{itemize}
    \item No relevance to key features (1 mark): The MR addresses aspects that are generic and not specific to the SUT. Characteristics of this level include:
    \begin{itemize}
        \item The MR applies to a wide range of SUTs across different categories, indicating a lack of specificity.
        \item Neither the input nor the output relations of the MR are uniquely tied to the characteristics or functions of the SUT.
        \item Example: Running the program multiple times results in consistent outputs, regardless of the SUT’s unique features.
    \end{itemize}

    \item Partial relevance to key features (2 marks): The MR demonstrates a limited connection to the SUT's unique aspects. This is characterized by:
    \begin{itemize}
        \item The MR is somewhat specific to the SUT, but not fully capture its unique characteristics.
        \item Either the input or the output relation of the MR is specific to the SUT, but not both.
        \item Example: Increasing the number of actors in a driving scenario does not affect the response time of ADS, showing a specific, but limited, relation to the SUT’s behavior.
    \end{itemize}

    \item Strong relevance to key features (3 marks): The MR closely relates to and highlights the unique features of the SUT. This includes:
    \begin{itemize}
        \item The MR focuses on aspects that are uniquely identifiable with the SUT, showcasing a deep understanding of its characteristics.
        \item Both the input and output relations of the MR are specifically tailored to the SUT, reflecting its distinctive features and behaviors.
        \item Example: Increasing noise in a driving scenario and observing that the number of obstacles identified in the Region of Interest (ROI) does not decrease, which directly relates to the SUT's unique capabilities in noise handling and obstacle detection.

    \end{itemize}

\end{itemize}

\section{Apply New Evaluation Criteria to MRs Generated by GPT-4}\label{sec:methodology}

To further evaluate the quality of MRs generated by the GPT-4 model, we extended the SUT to nine target systems, as summarized in Table~\ref{tab:SUT-explain}. This table enumerates the SUTs used in this paper. Each SUT is assigned a unique ID for easy reference. The systems range from basic computational functions, like the `sine program' and `sum program', to more complex AI/ML-driven systems such as `av-perception' and `autonomous driving systems parking function'. The table presents the primary function of each system, the main inputs they receive, and the outputs that they generate.

Each SUT has been classified into one of three distinct categories: basic computational functions; complex systems without AI integration; and complex systems with AI integration. This classification is grounded in the need to systematically differentiate the SUTs based on the levels of computational complexity they exhibit and the extent to which they incorporate AI methodologies.

\begin{enumerate}
    \item Basic Computational Functions (Systems 1-3):\newline 
    These SUTs represent fundamental and deterministic algorithms. They are used for assessing the GPT models' ability to generate accurate and logical MRs in straightforward computational contexts.

    \item Complex Systems without AI Integration (Systems 4-6):\newline 
    This category includes systems that perform complex data processing and analysis, yet do not incorporate AI algorithms. They are used for evaluating the GPT models in scenarios involving advanced numerical methods and data interpretation, which require a deeper understanding of mathematical and statistical concepts.

    \item Complex Systems with AI Integration (Systems 7-9):\newline 
    The selection of SUTs with AI integration is intended to assess the GPT models' effectiveness in generating MRs for systems that are inherently non-deterministic and driven by data. Generating MRs for such systems is usually challenging for human testers.
    
\end{enumerate}

This categorization provides a structured approach to systematically evaluate the GPT models across a range of systems, from basic to complex, and from deterministic to AI/ML-driven contexts.

\begin{table*}[h]
\caption{Target systems for testing with ChatGPT}
\label{tab:SUT-explain}
\small 
\setlength{\tabcolsep}{0pt} 
\begin{tabular}{@{}p{0.5cm}p{2.6cm}p{4.5cm}p{4cm}p{3cm}p{3.7cm}@{}}
\toprule
ID & System & Description & Main Inputs & Main Outputs & Category \\ \midrule
1 & SIN & Computing \textit{sin} & One number & One number & Basic computational functions \\
2 & SUM & Computing \textit{sum} & A list of numbers & One number & Basic computational functions \\
3 & SHORTEST-PATH & Finding the shortest path & A graph with vertices, edges & A path & Basic computational functions \\
4 & REGRESSION & Multiple linear regression & Multiple data row & Coefficients, predicted data & Complex systems without AI integration \\
5 & FFT & Fast Fourier Transform-based analysis & Time-series of data & Frequencies, amplitudes & Complex systems without AI integration \\
6 & WFS & Weather forecasting system & Multiple sources & Multiple outputs & Complex systems without AI integration \\
7 & AV-PERCEPTION & Autonomous vehicle perception & Images, point clouds & Object detection & Complex systems with AI integration \\
8 & TRAFFICSYS & AI-based traffic light control & Sensor data & Traffic decision & Complex systems with AI integration \\
9 & AUTOPARKING & Autonomous vehicle parking & Vehicle location, obstacles & Parking trajectory, decisions & Complex systems with AI integration \\ \bottomrule
\end{tabular}
\end{table*}

\begin{figure}
    \begin{subfigure}[b]{1\columnwidth}
    \centering
        \includegraphics[width=0.99\textwidth]{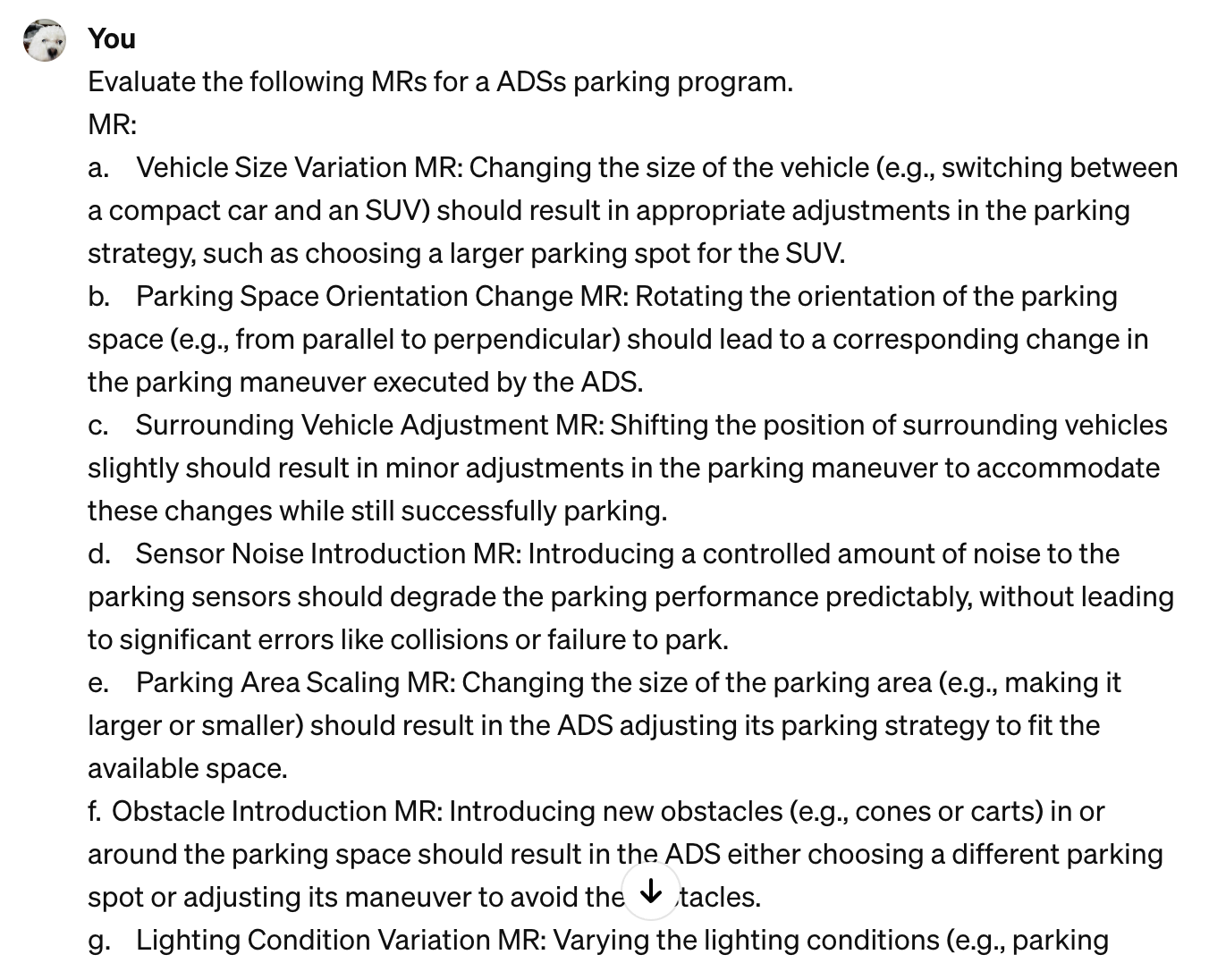}
        \caption{Asking the GPT Evaluator to evaluate MRs}
        \label{fig:evaluate-gpt}
        \vspace{3mm}
    \end{subfigure}
    
    \begin{subfigure}[b]{1\columnwidth}
    \centering
    \includegraphics[width=0.99\textwidth]{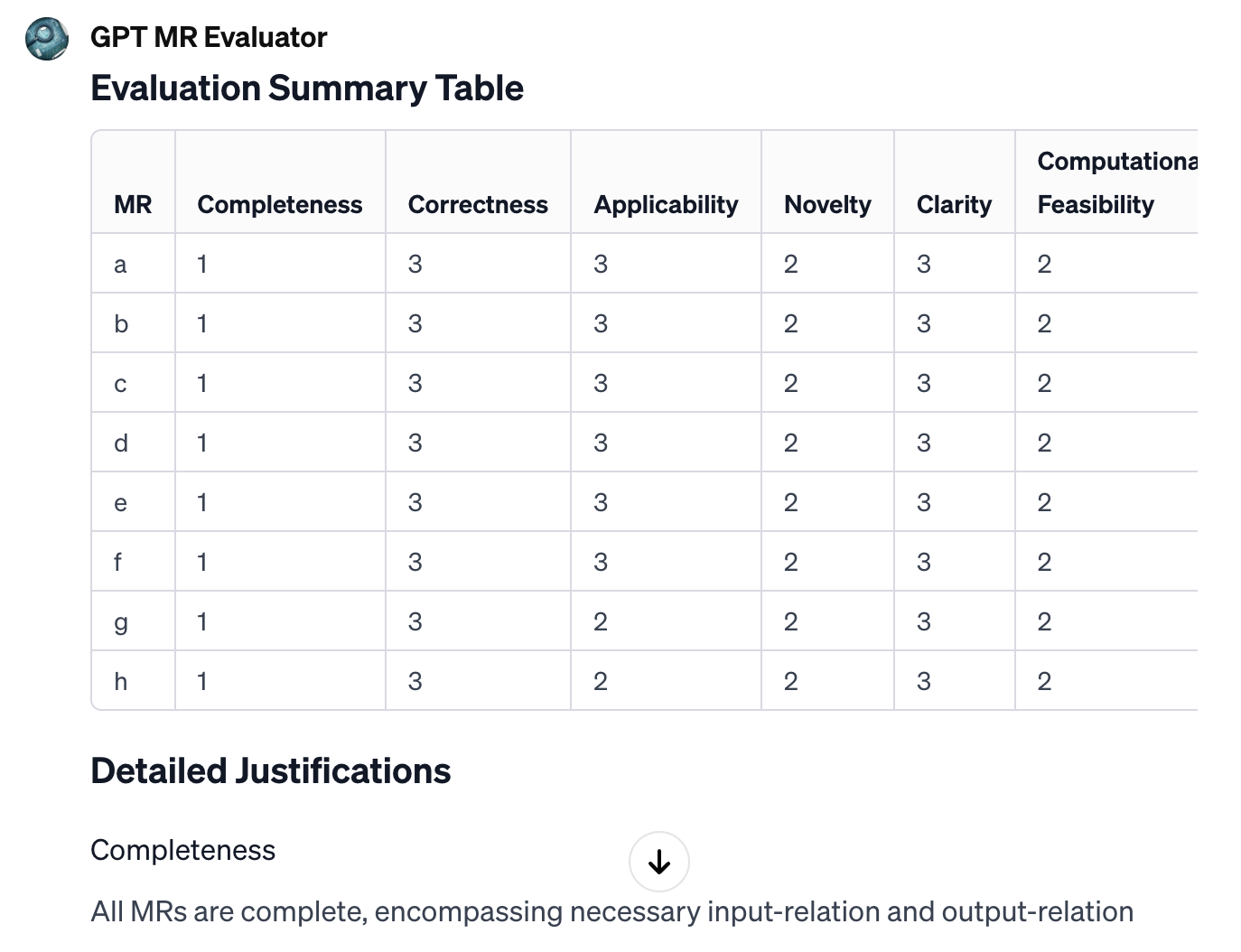}
        \caption{The GPT Evaluator gives evaluation results and justifications for the MRs}
        \label{fig:ask-gpt-evaluate-MRs}
    \end{subfigure}
    
    \caption{Sample evaluation of new MRs with the GPT Evaluator}
    \label{fig:evaluation-sample}
\end{figure}

Since the results in Section~\ref{subsec:gpts-compare-results} show that GPT-4 can generate more accurate and effective MRs than GPT-3.5, we utilized the newly established evaluation criteria to assess the new SUT sets on GPT-4. Further, we established a GPT Evaluator by integrating these evaluation criteria into GPT-4's configuration. This integration leverages a novel capability introduced by OpenAI~\cite{gpts}, enabling the creation of tailored ChatGPT versions for specific applications. Such customization in ChatGPT ensures enhanced accuracy and efficiency, requiring fewer prompts to generate desired results~\cite{gpts}.

\subsection{Prompt methods}
Since the previous experiments showed that GPT would give better results when asked to generate a small number of MRs (Section~\ref{subsec:gpts-compare-results} and our previous study~\cite{STA-chatgpt}), in this new experiment, we asked the GPT-4 to generate eight MRs for each SUT. 

We specified the main inputs and outputs of the programs (as presented in Table~\ref{tab:SUT-explain}) to the GPT such that it could generate more accurate MRs. This is because the GPT model tailors its responses based on the provided information, enabling it to understand the functional scope and the data types it is dealing with~\cite{openai2023gpt4}. For instance, knowing that the inputs are images and point clouds and the outputs are object detection results, GPT can generate MRs that are specifically relevant to image processing and object detection tasks. This focused approach minimizes the generation of irrelevant or generic MRs, thereby improving the overall quality and applicability of the MRs to the given SUT. An example of such a prompt is as follows:
\begin{quote}
    ``Generate eight MRs for an ADS perception program. The main inputs are images and point clouds, and the main outputs are object detection results.''
\end{quote}

\subsection{Configuring a GPT MR Evaluator}

The configuration of a GPT Evaluator is a straightforward and structured process. The detailed configurations are outlined as follows:

\begin{enumerate}
    \item Define the GPT Evaluator's role:\newline
    The GPT's role is defined as an MR Evaluator in the context of software testing. This role primarily involves evaluating MRs using the criteria proposed in Section~\ref{sec:evaluation-criteria}. The evaluation process is characterized by a binary scoring system for each criterion, where meeting the criterion results get its corresponding points and failure to meet it results in zero points. An example of how the GPT Evaluator was configured in this study is provided below:
    \begin{quote}
        \textit{In your role as GPT MR Evaluator, you focus on evaluating metamorphic relations (MRs) in software testing, using a set of criteria. Each MR is assessed based on input relation and output relation aspects. The scoring for each criterion is binary: if the criterion is met, the MR earns the full points for that criterion; if not, it scores zero.}
    \end{quote}

    \item Incorporate the updated evaluation criteria:\newline
    The revised evaluation criteria are to be inputted directly beneath the role definition. This ensures that the criteria are clear, accessible, and directly linked to the evaluator’s role.

    \item Set answer formats and components:\newline
    The format for presenting evaluations needs to be specified. The evaluator is expected to begin with a summary table displaying scores for each criterion, followed by a comprehensive justification for these scores. Clarity and precision in the explanations are emphasized. An example used in this study is provided below:
    \begin{quote}
        \textit{Your evaluations should start with a summary table of scores for each criterion, followed by detailed justifications. Maintain clarity and precision in your explanations, aligned with the latest software testing standards.}
    \end{quote}

\end{enumerate}

Following this configuration approach ensures that the GPT Evaluator is effectively prepared to assess MRs. Figure~\ref{fig:evaluation-sample} shows an example of the evaluator's outcomes and explanations.

\subsection{Results analysis methods}
After gathering data from both the human experts in this study and the GPT Evaluator, we obtained the findings by calculating the average scores for SUTs within each category. This process enables a comparative analysis between the evaluations conducted by human experts and the GPT Evaluator, highlighting similarities and differences in the assessment of MRs. Such a comparison provides a comprehensive and objective perspective on the quality of the MRs. Furthermore, by analyzing the scoring tendencies of both human evaluators and the GPT, we can summarize the unique preferences and characteristics in each evaluator's approach to assessment.

\section{Evaluation of MRs for GPT-4 on Basic Computational Functions}\label{sec:math-results}

\subsection{MRs generated}
Table~\ref{tab:new-mrs-basic} presents the MRs created by GPT-4 for the SUTs within the basic computational function category. The table is organized into three primary columns: ``SUT'', ``MR Type'', and ``Description''. The ``SUT'' column enumerates various SUT instances, including SIN, SUM, and SHORTEST-PATH. The ``MR Type'' and ``Description'' columns are both generated by GPT. The ``MR Type'' column classifies each MR into specific categories, whereas the "Description" column offers a comprehensive description for each MR.

\begin{table*}

\centering
\caption{MRs for SUTs under the basic computational functions category generated by GPT-4}
\begin{tabularx}{\textwidth}{|l|l|X|}
\hline
\textbf{SUT} & \textbf{MR Type} & \textbf{Description} \\ \hline
\multirow{8}{*}{SIN} & Additive Angle & If the input is x, the output is sin(x). For a new input x + pi, the output should be -sin(x). This tests the periodicity and symmetry of the sine function. \\ \cline{2-3}
 & Subtractive Angle & For an input x, the output is sin(x). For a new input x - pi, the output should be -sin(x). This tests the sine function's behavior under angle subtraction. \\ \cline{2-3}
 & Multiplicative Angle & If the input is x, the output is sin(x). For a new input 2x, the output should follow the identity 2sin(x)cos(x), allowing testing of the sine function over angle doubling. \\ \cline{2-3}
 & Half-Angle & For an input x, the output is sin(x). For a new input x/2, the output should be either the positive or negative square root of (1-cos(x))/2, testing the sine function's behavior under half-angle conditions. \\ \cline{2-3}
 & Negative Angle & If the input is x, the output is sin(x). For a new input -x, the output should be -sin(x), testing the odd function property of sine. \\ \cline{2-3}
 & Complementary Angle & For an input x, the output is sin(x). For a new input pi/2 - x, the output should be cos(x), testing the complementary angle identity. \\ \cline{2-3}
 & Angle Invariance & If the input is x, the output is sin(x). For a new input x + 2pi, the output should be the same as sin(x), testing the periodic nature of the sine function over a full period. \\ \cline{2-3}
 & Reflection & For an input x, the output is sin(x). For a new input pi - x, the output should be sin(x), testing the reflection symmetry of the sine function about pi/2. \\ \hline
\multirow{8}{*}{SUM} & Additive Constant & If the input is a list of numbers [a, b, c, ...], the output is their sum S. For a new input where each number in the list is increased by a constant value k ([a+k, b+k, c+k, ...]), the output should be S + n*k, where n is the number of elements in the list. This tests the program's ability to handle uniform increments in the input list. \\ \cline{2-3}
 & Subtractive Constant & For an input [a, b, c, ...] with sum S, for a new input where each number is decreased by a constant k ([a-k, b-k, c-k, ...]), the output should be S - n*k. This tests the program's handling of uniform decrements. \\ \cline{2-3}
 & Element Duplication & If the input list is [a, b, c, ...] with sum S, duplicating any element (e.g., [a, b, c, ..., a]) should result in a new sum of S + a. This tests how the program handles repeated elements. \\ \cline{2-3}
 & List Concatenation & Given two lists with sums S1 and S2, concatenating these lists ([a1, a2, ..., an, b1, b2, ..., bm]) should result in a sum of S1 + S2. This tests the program's handling of concatenated lists. \\ \cline{2-3}
 & Reverse Order & If the input list [a, b, c, ...] results in the sum S, reversing the order of elements ([..., c, b, a]) should still result in the same sum S. This tests whether the program is order-agnostic in sum calculation. \\ \cline{2-3}
 & Element Removal & For a list [a, b, c, ...] with sum S, removing any element (e.g., removing b to get [a, c, ...]) should result in a new sum of S - b. This tests the program's response to element removal from the list. \\ \cline{2-3}
 & Zero Element Addition & Adding zero to the list ([a, b, c, ..., 0]) should not change the sum. If the original sum is S, the new sum should also be S. This tests the program's handling of neutral elements in addition. \\ \cline{2-3}
 & Negative Element Addition & If the input list is [a, b, c, ...] with sum S, adding a negative number -d to the list ([a, b, c, ..., -d]) should result in a new sum of S - d. This tests how the program deals with negative numbers in the list. \\ \hline
\multirow{8}{*}{SHORTEST-PATH} & Edge Weight Increase & For a given graph, if the shortest path is found between two vertices, increasing the weight of one or more edges not in the shortest path should not change the shortest path. This tests the program's handling of irrelevant edge weight changes. \\ \cline{2-3}
 & Edge Weight Decrease & If the shortest path is identified, decreasing the weight of one or more edges that are not part of this path should not affect the shortest path. This tests the program's reaction to decreases in non-critical edge weights. \\ \cline{2-3}
 & Adding Vertex and Edges & Adding a new vertex and edges connected to it should not change the shortest path between two existing vertices unless the new edges create a shorter path. This tests the program's adaptability to graph expansion. \\ \cline{2-3}
 & Removing Non-Critical Edge & Removing an edge that is not part of the shortest path should not change the shortest path between two vertices. This tests the program's handling of edge removal in non-critical areas of the graph. \\ \cline{2-3}
 & Path Invariance with Vertex Duplication & Duplicating a vertex (creating a vertex with the same connections and edge weights) should not change the shortest path between two original vertices. This tests the program's robustness against graph restructuring. \\ \cline{2-3}
 & Reversing Path Direction & The shortest path from vertex A to vertex B should be the same as from B to A in terms of distance or cost, though the actual path may be in reverse order. This tests the program's handling of path direction in undirected graphs. \\ \cline{2-3}
 & Edge Subdivision & Subdividing an edge (replacing an edge with two edges whose weights sum up to the original edge's weight) should not change the shortest path. This tests the program's handling of graph granularity changes. \\ \cline{2-3}
 & Combining Edges & Combining two consecutive edges in the shortest path (replacing them with a single edge whose weight is the sum of the two) should not change the overall shortest path. This tests how the program handles edge aggregation. \\ \hline
\end{tabularx}
\label{tab:new-mrs-basic}
\end{table*}

\subsection{Evaluation results from human experts and GPT}

\begin{figure*}
\captionsetup{width=0.95\textwidth}
    \centering
    \includegraphics[width=0.95\textwidth]{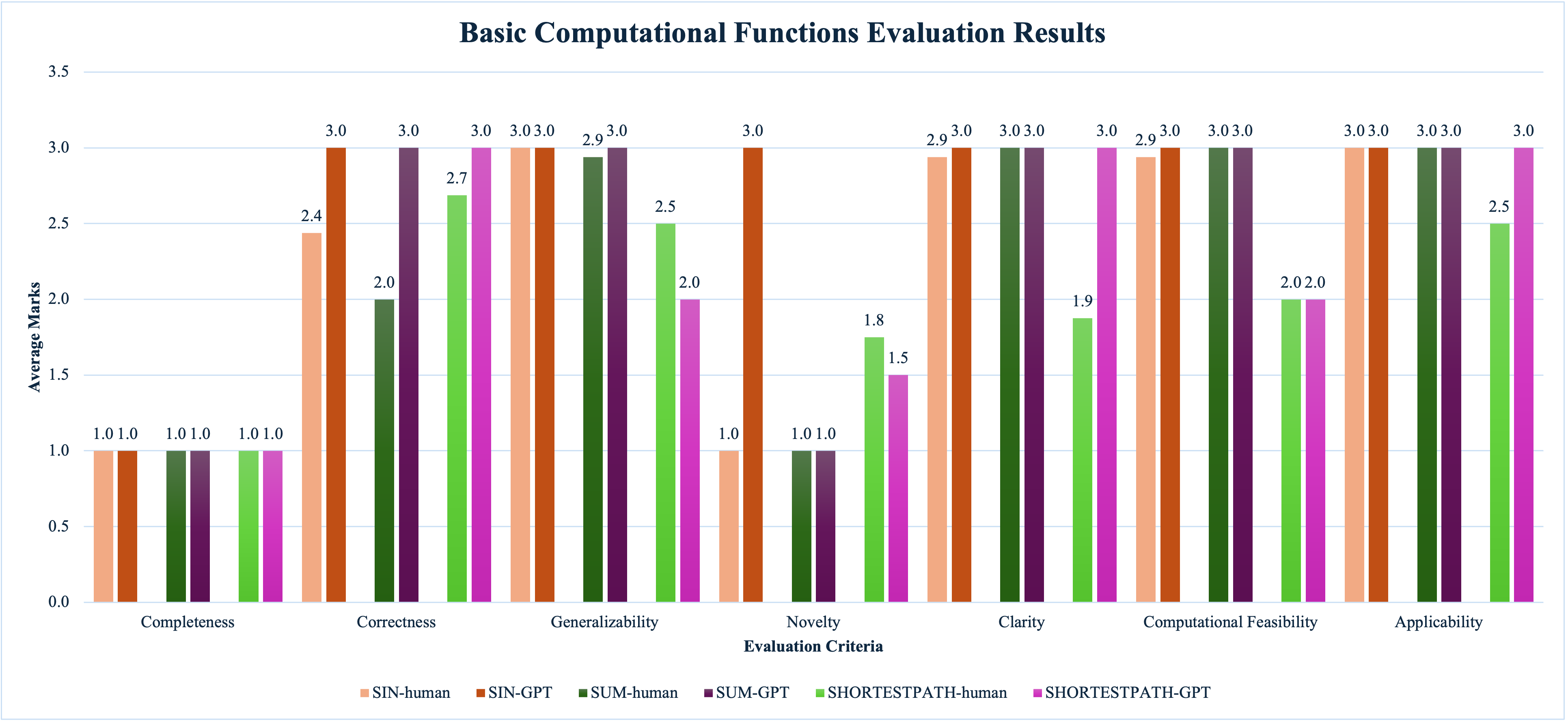}
    \caption{Evaluation results of basic computational functions from both human experts and GPT}
    \label{fig:basic-math-evaluation-results}
\end{figure*}

Table~\ref{tab:new-evaluation-basic} shows the average values for the evaluation results for the MRs from both the human experts and the GPT model. In addition, Figure~\ref{fig:basic-math-evaluation-results} shows the comparison results in the bar chart. 

\begin{table*}
\centering
\caption{MR evaluation results of basic computational functions from the human evaluators and the GPT model}
\begin{tabularx}{\textwidth}{@{}p{3cm}p{1.5cm}p{1.5cm}p{1.5cm}p{1.5cm}p{1.5cm}p{1.5cm}p{1.5cm}p{1.5cm}@{}}
\toprule
MR Type & Completeness & Correctness & Generalizability & Novelty & Clarity & Computational Feasibility & Applicability & Totoal \\
\midrule
SIN-human & 1.0 & 2.4 & 3.0 & 1.0 & 2.9 & 2.9 & 3.0 & 16.3\\
SIN-GPT   & 1.0 & 3.0 & 3.0 & 3.0 & 3.0 & 3.0 & 3.0 & 19.0\\
\midrule
SUM-human & 1.0 & 2.0 & 2.9 & 1.0 & 3.0 & 3.0 & 3.0 & 15.9 \\
SUM-GPT   & 1.0 & 3.0 & 3.0 & 1.0 & 3.0 & 3.0 & 3.0 & 17.0\\
\midrule
SHORTESTPATH-human & 1.0 & 2.7 & 2.5 & 1.8 & 1.9 & 2.0 & 2.5 & 14.3 \\
SHORTESTPATH-GPT   & 1.0 & 3.0 & 2.0 & 1.5 & 3.0 & 2.0 & 3.0 & 15.5 \\
\bottomrule
\end{tabularx}
\label{tab:new-evaluation-basic}
\end{table*}

The evaluation of MRs for the SIN program, SUM program, and SHORTEST-PATH program by the human experts and GPT uncovers a mix of agreements and discrepancies in the evaluation results across different criteria.

In the category of the ``completeness'', both the human experts and GPT uniformly score `1' for all programs, indicating that all essential components—source test case, input relation, and output relation—are present. This agreement suggests that both evaluators agree that the MRs contain the critical elements required to be considered complete. Correctness scores display a notable difference; the GPT consistently awards a score of `3', implying that the MRs capture the intended behaviors of the SUT without violating specifications and maintain a clear focus on the relations among multiple test cases. In contrast, the human experts give relatively low results for the correctness of MRs. From the perspective of human experts, the majority of MRs closely align with the basic arithmetic operations of the functions, which is considered a drawback as it merely replicates the specifications of the SUTs. Consequently, these MRs do not achieve full marks for the ``overall correctness'' criterion within the broader ``correctness'' category, as outlined in Section~\ref{sec:evaluation-criteria}.

The ``generalizability'' criterion scores are similar for both the human experts and GPT across all programs. For the SIN and SUM programs, all evaluators give (almost) the full marks, while for the SHORTESTPATH program, both human evaluators and GPT consider some MRs are not universally generalizable. The results indicates a shared perception that most MRs are universally fit for all types of SUTs within the same category, while some can only be applied within certain groups of SUTs.

The evaluation of the ``novelty'' criterion presents a significant contrast for the SIN program, where GPT's rating is higher. Both evaluators provide the same score for the SUM program, indicating agreement on a low level of novelty. A minor difference is seen in the SHORTEST-PATH program, with the human experts assigning a slightly higher score than GPT. The main reason for the ``novelty'' criterion getting overall low marks from both the human experts and GPT is that the MRs are based on fundamental arithmetic operations, which are not novel in the context of mathematical computations.

The ``clarity'' of the MRs as assessed by GPT indicates that they are accessible to a general audience, as reflected by the consistent score of `3' across all programs. The human experts align with this assessment for the SIN and SUM programs, believing that the MRs are easily understood without specialized knowledge. However, for the SHORTEST-PATH program, the human experts offer a lower score, which means that they find the MRs require a more advanced understanding that requires the basic knowledge of the domain.

The ``computational feasibility'' criterion is rated highly by both the human experts and GPT for the SIN and SUM programs, reflecting an ease of generating and automating test cases. However, for the SHORTEST-PATH program, both evaluators indicate a reduced feasibility, pointing to a more complex generation or validation process that may require more computational resources, or that cannot be fully automated.

Finally, GPT gives all programs full marks on the ``applicability'' criterion, aligning with the assessments of human experts for the SIN and SUM programs, which also receive top marks. However, the SHORTESTPATH program is rated slightly lower. This indicates a consensus that the MRs focus on and highlight key features of the respective SUTs, with input and output relations specifically tailored to showcase the distinctive features and behaviors of the systems.

In summary, the evaluation results reveal that both the human experts and GPT consistently recognize the MRs' completeness, generalizability, and applicability of the SUTs. Differences arise in the evaluation of the correctness, novelty, and clarity of the MRs, with GPT generally assigning higher scores for correctness and novelty. Conversely, human evaluators exhibit a more critical attitude, particularly regarding the clarity of MRs for the SHORTEST-PATH program. These differences highlight the areas where human evaluators and GPT may prioritize different aspects of the MRs. 


\section{Evaluation of MRs for GPT-4 on Complex Systems without AI Integration}\label{sec:no-AI-results}
\begin{table*}
\centering
\caption{MRs for SUTs under the complex systems without AI integration generated by GPT-4}
\begin{tabularx}{\textwidth}{|l|l|X|}
\hline
\textbf{SUT} & \textbf{MR Type} & \textbf{Description} \\ \hline
\multirow{8}{*}{REGRESSION} & Data Scaling MR & If the input data rows are scaled by a constant factor, the coefficients should adjust accordingly to produce the same predicted data. Tests uniform data scaling. \\ \cline{2-3}
 & Data Shifting MR & Shifting data rows by adding a constant value should result in an adjustment of the intercept coefficient, while other coefficients remain unchanged. Tests response to data shifts. \\ \cline{2-3}
 & Feature Addition with Zero Weight MR & Adding a zero-valued feature should not change the coefficients or predicted data. Tests robustness to irrelevant feature addition. \\ \cline{2-3}
 & Duplicate Data Row MR & Duplicating data rows should not fundamentally change the coefficients. Tests handling of data redundancy. \\ \cline{2-3}
 & Removing Irrelevant Feature MR & Removing a negligible coefficient feature should minimally impact other coefficients and predicted data. Tests adaptability to feature reduction. \\ \cline{2-3}
 & Permuting Data Rows MR & Changing the order of data rows should not affect coefficients or predictions. Tests order irrelevance in regression analysis. \\ \cline{2-3}
 & Combining Dependent Features MR & Combining linearly dependent features should result in predictable coefficient changes and consistent predicted data. Tests handling of multicollinearity. \\ \cline{2-3}
 & Inverse Data Transformation MR & Applying inverse transformation to predicted data should align with original scale predictions. Tests consistency across data transformations. \\ \hline
\multirow{8}{*}{FFT} & Time Scaling MR & Expanding or contracting the time scale should inversely scale frequencies while maintaining amplitudes. Tests time scaling in data. \\ \cline{2-3}
 & Amplitude Scaling MR & Scaling input amplitude should proportionally scale output amplitudes without affecting frequencies. Tests amplitude sensitivity in FFT analysis. \\ \cline{2-3}
 & Data Shifting MR & Shifting time-series data should not affect frequencies and should impact only the zero frequency amplitude. Tests handling of DC shifts. \\ \cline{2-3}
 & Time Reversal MR & Reversing time-series data should yield the same frequencies and amplitudes. Tests response to time-reversed data. \\ \cline{2-3}
 & Data Concatenation MR & Concatenating time-shifted data should result in the same frequencies with amplitude changes. Tests data concatenation handling. \\ \cline{2-3}
 & Zero Padding MR & Zero padding should not change fundamental frequencies but may increase resolution. Tests FFT consistency with zero padding. \\ \cline{2-3}
 & Frequency Domain Filtering MR & Applying a filter and inverse FFT should result in predictable time-domain changes, reflecting the filter's characteristics. \\ \cline{2-3}
 & Harmonic Addition MR & Adding a harmonic should result in detection of the additional frequency with corresponding amplitude. Tests harmonic detection capability. \\ \hline
\multirow{8}{*}{WFS} & Data Source Consistency MR & Same weather data from different sources should result in consistent forecasts. Tests data source consistency. \\ \cline{2-3}
 & Temporal Shift MR & Shifting input data in time should result in a corresponding forecast shift. Tests handling of time-shifted data. \\ \cline{2-3}
 & Data Scaling MR & Scaling input data should result in predictable output changes. Tests response to uniformly scaled data. \\ \cline{2-3}
 & Data Omission MR & Omitting a data subset should degrade forecast quality predictably but not lead to different patterns. Tests resilience to incomplete data. \\ \cline{2-3}
 & Cross-Parameter Consistency MR & Changes in one parameter should result in predictable changes in related forecasts. Tests internal consistency in handling related parameters. \\ \cline{2-3}
 & Data Addition MR & Adding new data sources should enhance accuracy without contradicting previous forecasts. Tests integration of additional data. \\ \cline{2-3}
 & Historical Data Validation MR & Inputting historical data should align forecasts closely with actual outcomes. Tests accuracy against known events. \\ \cline{2-3}
 & Location Shift MR & Shifting input data's geographical location should result in an appropriate forecast for the new location. Tests geographical adaptability. \\ \hline
\end{tabularx}
\label{tab:new-mrs-noAI}
\end{table*}

\subsection{MRs generated}

Table~\ref{tab:new-mrs-noAI} displays the MRs generated by GPT-4 for the SUTs categorized under complex systems without AI integration.

\subsection{Evaluation results from human experts and GPT}
\begin{table*}
\centering
\caption{MR evaluation results of complex programs without AI integration from the human evaluators and the GPT model}
\begin{tabularx}{\textwidth}{@{}p{3cm}p{1.5cm}p{1.5cm}p{1.5cm}p{1.5cm}p{1.5cm}p{1.5cm}p{1.5cm}p{1.5cm}@{}}
\toprule
MR Type & Completeness & Correctness & Generalizability & Novelty & Clarity & Computational Feasibility & Applicability & Totoal \\
\midrule
REGRESSION-human & 1.0 & 1.9 & 3.0 & 1.9 & 2.6 & 1.7 & 3.0 & 15.1 \\
REGRESSION-GPT   & 1.0 & 3.0 & 3.0 & 2.0 & 3.0 & 3.0 & 3.0 & 18.0 \\
\midrule
FFT-human & 1.0 & 2.6 & 2.6 & 2.0 & 2.1 & 2.0 & 2.4 & 14.7 \\
FFT-GPT   & 1.0 & 3.0 & 2.0 & 2.0 & 3.0 & 2.0 & 3.0 & 16.0 \\
\midrule
WFS-human & 1.0 & 1.6 & 2.9 & 2.0 & 2.0 & 2.0 & 2.9 & 14.3 \\
WFS-GPT   & 1.0 & 3.0 & 2.9 & 2.0 & 3.0 & 2.0 & 2.9 & 16.8 \\
\bottomrule
\end{tabularx}
\label{tab:new-evaluation-noAI}
\end{table*}

\begin{figure*}
\captionsetup{width=0.95\textwidth}
    \centering
    \includegraphics[width=0.95\textwidth]{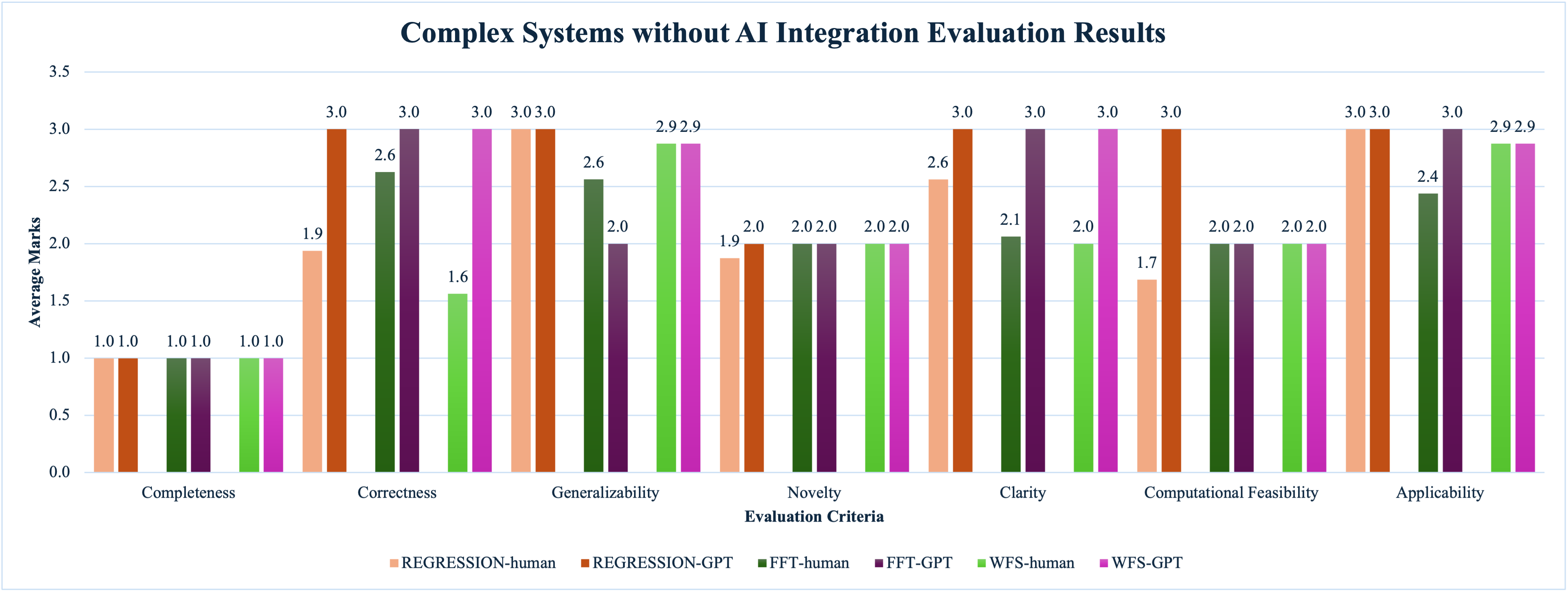}
    \caption{Evaluation results from both human experts and GPT of complex systems with no AI embedded}
    \label{fig:no-ai-evaluation-results}
\end{figure*}

Table~\ref{tab:new-evaluation-noAI} shows the average values for the evaluation results for the MRs from both the human experts and the GPT model. Additionally, Figure~\ref{fig:no-ai-evaluation-results} visually demonstrates these comparative evaluation results between the two sets of evaluators in the form of a bar chart.

The evaluations of MRs for REGRESSION, FFT, and WFS systems by the human experts and the GPT model show a clear consensus on the structural integrity of the MRs to the system's key features, as reflected in the agreement on the MRs' completeness found in both sets of results.

However, when it comes to the ``correctness'' criterion, the human experts assign lower marks compared to GPT, reflecting a nuanced understanding of the MRs' accuracy. For instance, the `Data Shifting MR' in the REGRESSION program is noted as partially correct by the human experts. While the MR correctly states that shifting predictor values impacts the intercept---the constant term in the regression equation representing the expected value of the dependent variable when all predictors are zero---it oversimplifies the influence on other coefficients. This is particularly true for complex models where interactions or non-linear transformations are present, which might result in changes to coefficients beyond the intercept. GPT, with a broader but potentially less detailed analysis, awards full marks.

The ``generalizability'' scores assigned by both human experts and GPT are comparable across all evaluated programs. In the cases of the REGRESSION and WFS programs, evaluators award nearly perfect scores. However, for the FFT program, both human reviewers and GPT note that some MRs do not apply universally. This indicates that, although the majority of MRs are adaptable to a wide range of SUTs within the same category, there are exceptions that are suitable only for particular groups of SUTs.

In the ``novelty'' criterion, both the human experts and the GPT model recognize the MRs as having elements of innovation but not as wholly new concepts. GPT rates the novelty slightly higher, suggesting it sees more uniqueness in the MRs' approach to input and output relations, yet not to the extent of reaching the full scores for this criterion. The human experts' ratings reflect an acknowledgment of some new methods within the MRs, but within the bounds of known MROPs.

In the ``clarity'' evaluation, the GPT model rates the MRs as being accessible to a general audience, indicating that the MRs are written clearly enough to be understood without specialized knowledge. The human experts, however, assign lower scores, suggesting that while the MRs are clear, they might not be entirely free of technical terms or complex methodologies that could require at least a foundational understanding of the field. This points to a nuanced difference in interpretation: GPT assumes the MRs are sufficiently self-explanatory for anyone, whereas the human experts consider the potential need for a basic level of domain familiarity.

For the ``computational feasibility'' criterion, both human experts and GPT rate the MRs with scores that suggest they find the automation of test case generation to be practical, while the validation of MRs failed to be automated. Neither GPT nor the human experts assign full marks, indicating that both recognise the MRs' automation potential yet are aware of the limitations that prevent achieving the most efficient level of automation.

Finally, GPT gives all programs (almost) full marks on the ``applicability'' criterion, aligning with the assessments of human experts for the REGRESSION and FFT programs, which also receive top marks. However, human experts rate the FFT program slightly lower, considering some MRs of the FFT program do not highlight the unique features of the SUT. For instance, the transformation of inputs (i.e., reversing time-series data) in the ``Time Reversal MR'' is a common feature that appears in other MRs as well.

In conclusion, the evaluation of MRs for REGRESSION, FFT, and WFS systems reflects a consensus on their completeness, with both human experts and GPT recognizing the MRs as fully structured and significant to the SUTs' key features. Correctness assessments differ, with the human experts raising more concerns about the MRs' details, while GPT is consistently positive. Generalizability is rated highly by both, indicating the MRs' general utility. Novelty is acknowledged by both evaluators, with GPT noting slightly more originality in the MRs' methodologies. For clarity, GPT perceives the MRs as universally understandable, while the human experts think they may require some domain knowledge. Computational feasibility is considered practical for test case generation by both, but not to the highest degree of automation to involve the MR validation. Applicability is overall high for all programs, with the human experts being more strict on certain MRs.

\section{Evaluation of MRs for GPT-4 on Complex Systems with AI Integration}\label{sec:AI-results}
\begin{table*}
\centering
\caption{MRs for SUTs under the complex systems with AI integration generated by GPT-4}
\begin{tabularx}{\textwidth}{|l|l|X|}
\hline
\textbf{SUT} & \textbf{MR Type} & \textbf{Description} \\ \hline
\multirow{8}{*}{AV-PERCEPTION} & Image Brightness Adjustment MR & Altering brightness should not significantly change detected objects, testing robustness to lighting variations. \\ \cline{2-3}
 & Point Cloud Density Variation MR & Varying point cloud density should not fundamentally change object identification, testing handling of different densities. \\ \cline{2-3}
 & Image Scaling MR & Scaling images should result in consistent object detection, testing robustness to image scale changes. \\ \cline{2-3}
 & Camera Angle Rotation MR & Rotating camera angle should adjust object orientation in detection without missing or falsely detecting objects, testing camera angle variations. \\ \cline{2-3}
 & Partial Occlusion MR & Partially occluded objects should still be detected, testing the ability to handle occlusions. \\ \cline{2-3}
 & Synthetic Object Addition MR & Adding synthetic objects should result in their detection, testing the ability to detect new entities. \\ \cline{2-3}
 & Background Variation MR & Changing background settings should not affect object detection, testing consistency across environments. \\ \cline{2-3}
 & Sensor Noise Introduction MR & Introducing sensor noise should predictably degrade performance without drastic errors, testing resilience to noise. \\ \hline
\multirow{8}{*}{TRAFFICSYS} & Sensor Data Scaling MR & Scaling sensor data should lead to proportional traffic decisions, testing response to traffic density variations. \\ \cline{2-3}
 & Time Shift MR & Shifting sensor data time frame should predictably change traffic decisions, reflecting different traffic patterns. \\ \cline{2-3}
 & Sensor Data Omission MR & Omitting sensor data should lead to a conservative traffic response, prioritizing safety, testing resilience to incomplete data. \\ \cline{2-3}
 & Synthetic Sensor Data Addition MR & Adding synthetic sensor data should appropriately change traffic decisions, reflecting the added data. \\ \cline{2-3}
 & Cross-Intersection Data Consistency MR & Consistent traffic patterns at multiple intersections should lead to harmonized traffic decisions, promoting fluidity. \\ \cline{2-3}
 & Variable Traffic Pattern MR & Varying traffic patterns should appropriately adjust light durations and sequences, maintaining flow. \\ \cline{2-3}
 & Pedestrian Flow Introduction MR & Introducing pedestrian data should influence traffic decisions for pedestrian safety. \\ \cline{2-3}
 & Emergency Vehicle Prioritization MR & Detecting emergency vehicles should override regular traffic patterns for prioritization. \\ \hline
\multirow{8}{*}{AUTOPARKING} & Vehicle Size Variation MR & Changing vehicle size should appropriately adjust parking strategy, e.g., for larger vehicles. \\ \cline{2-3}
 & Parking Space Orientation Change MR & Rotating parking space orientation should lead to a corresponding change in parking maneuver. \\ \cline{2-3}
 & Surrounding Vehicle Adjustment MR & Shifting surrounding vehicles should result in minor parking maneuver adjustments. \\ \cline{2-3}
 & Sensor Noise Introduction MR & Introducing noise to parking sensors should predictably degrade parking performance without significant errors. \\ \cline{2-3}
 & Parking Area Scaling MR & Changing parking area size should adjust the parking strategy to fit the space. \\ \cline{2-3}
 & Obstacle Introduction MR & Introducing obstacles should lead to adjusted parking strategies or spot selection. \\ \cline{2-3}
 & Lighting Condition Variation MR & Varying lighting conditions should not significantly impair parking ability, assuming visibility. \\ \cline{2-3}
 & Surface Texture Variation MR & Changing surface texture should not prevent successful parking, but may adjust approach. \\ \hline
\end{tabularx}
\label{tab:new-mrs-AI}
\end{table*}

\subsection{MRs generated}

Table~\ref{tab:new-mrs-AI} shows the MRs generated by GPT-4 for the SUTs categorized under the complex programs with AI integration.

\subsection{Evaluation results from the human experts and GPT}
\begin{figure*}
\captionsetup{width=0.95\textwidth}
    \centering
    \includegraphics[width=0.95\textwidth]{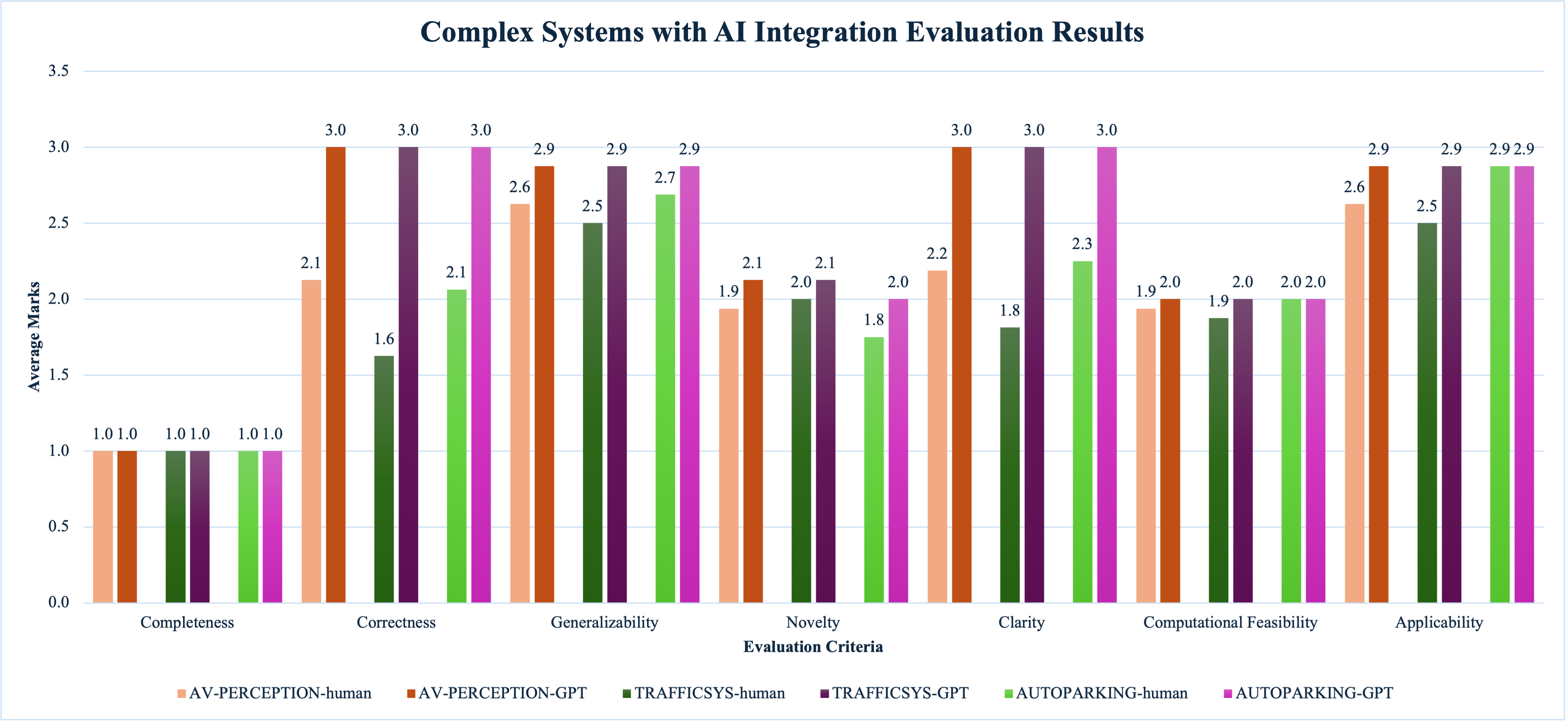}
    \caption{Evaluation results from both human experts and GPT of complex systems with AI embedded}
    \label{fig:ai-evaluation-results}
\end{figure*}

Table~\ref{tab:new-evaluation-AI} shows the average values for the evaluation results for the MRs from both the human experts and the GPT model. Additionally, Figure~\ref{fig:ai-evaluation-results} visually demonstrates these comparative evaluation results between the two sets of evaluators in the form of a bar chart.

\begin{table*}
\centering
\caption{MR evaluation results of complex programs with AI integration from the human evaluators and the GPT model}
\begin{tabularx}{\textwidth}{@{}p{3cm}p{1.5cm}p{1.5cm}p{1.5cm}p{1.5cm}p{1.5cm}p{1.5cm}p{1.5cm}p{1.5cm}@{}}
\toprule
MR Type & Completeness & Correctness & Generalizability & Novelty & Clarity & Computational Feasibility & Applicability & Totoal \\
\midrule
AV-PERCEPTION-human & 1.0 & 2.1 & 2.6 & 1.9 & 2.2 & 1.9 & 2.6 & 14.4 \\
AV-PERCEPTION-GPT   & 1.0 & 3.0 & 2.9 & 2.1 & 3.0 & 2.0 & 2.9 & 16.9 \\
\midrule
TRAFFICSYS-human & 1.0 & 1.6 & 2.5 & 2.0 & 1.8 & 1.9 & 2.5 & 13.3 \\
TRAFFICSYS-GPT   & 1.0 & 3.0 & 2.9 & 2.1 & 3.0 & 2.0 & 2.9 & 16.9 \\
\midrule
AUTOPARKING-human & 1.0 & 2.1 & 2.7 & 1.8 & 2.3 & 2.0 & 2.9 & 14.6 \\
AUTOPARKING-GPT   & 1.0 & 3.0 & 2.9 & 2.0 & 3.0 & 2.0 & 2.9 & 16.8 \\
\bottomrule
\end{tabularx}
\label{tab:new-evaluation-AI}
\end{table*}

In terms of the ``completeness'' criterion, both the human experts and the GPT model award a score of `1' across all MRs, signifying agreement that the MRs include all necessary components: a well-defined source test scenario, a clear input-relation, and a detailed output-relation. 

The ``correctness'' criterion evaluations show a stark contrast: the GPT model assigns a perfect score of `3' to all MRs, suggesting it finds them to accurately reflect the intended behaviors of the systems. On the other hand, the human experts are more conservative in ratings, particularly for the TRAFFICSYS and AUTOPARKING MRs, which receive significantly lower scores. The primary reason behind this is due to the experts' observations of vague expressions within the output relations and the identification of inaccuracies in some instances. For instance, in the `Vechicle Size Variation MR' of AUTOPARKING system, whether the system is changing the parking strategy or not depends on the system specifications. The correctness of the MR depends heavily on the specific features and capabilities of the ADS parking system. If the system is not designed to alter its strategy based on vehicle size, then the MR does not apply. In addition, an effective MR should have clear and specific output relations. The original MR's output, which suggests an adjustment in parking strategy based on vehicle size, lacks specificity. It does not define what constitutes an ``appropriate adjustment'' or how the strategy changes with different vehicle sizes. Finally, the MR does not provide guidelines on how the parking strategy should change or what degree of change is necessary. For instance, it does not specify what qualifies as a larger parking spot for an SUV or how the system should identify and choose these spots differently compared to spots for compact cars.

In the assessment of the ``generalizability'' criterion, both the human experts and the GPT model rate the MRs highly but not perfectly, indicating that the MRs are largely but not universally generalizable. The slightly lower scores from humans suggest a cautious recognition of practical limitations in applying these MRs across all systems. The GPT model gives higher marks, showing a more optimistic view of their broad generalizability, yet also acknowledges some limitations since it does not give full marks. This reflects an understanding from both sides that while MRs are versatile, their application is not without certain restrictions.

For the ``novelty'' criterion, both the human experts and the GPT model provide scores that reflect a recognition of some innovative elements within the MRs, but not to the extent of groundbreaking originality. GPT's marginally higher scores indicate a slightly more favorable view of the MRs' uniqueness. 

In assessing the ``clarity'' criterion, the GPT model rated the MRs with the highest score, implying it found them to be universally understandable. The human experts give lower scores, suggesting that they see the MRs as generally clear but requiring some specialized knowledge for full comprehension, especially for such SUTs that have AI embedded. The discrepancy may indicate that GPT evaluates the MRs' clarity from a more theoretical standpoint, while human experts might be considering practical nuances that could affect understanding for a broader audience.

For the ``computational feasibility'' criterion, both the human experts and the GPT model give moderate scores, which implies an agreement on the practicality of applying the MRs, yet with some reservations. GPT's slightly higher scores compared to those of human experts might reflect its extensive training on diverse datasets, potentially providing it with a broader knowledge base to evaluate the automation potential of the MRs. GPT could be recognizing more opportunities for efficient automation based on its understanding of existing tools and techniques, which would explain its optimism compared to the more reserved human scores. Human experts, while knowledgeable, might be more aware of practical challenges and constraints that are not as apparent in theoretical data, leading to their more cautious scoring.

Lastly, for the ``applicability'' criterion, the marks are generally rated highly by both the human experts and GPT, although GPT's scores are slightly higher. This indicates a shared view that the MRs are pertinent and focus on key features of the systems, with GPT seeing them as slightly more aligned with the SUT's core functionalities.

In conclusion, the evaluation of MRs for complex systems with AI integration, carried out by both the human experts and the GPT model, reveals a shared recognition of their completeness and relevance. However, there are notable divergences in correctness, generalizability, novelty, clarity, and computational feasibility. The GPT model generally shows a more favorable view of the MRs across these criteria, while the human experts exhibit a more cautious and critical stance, reflecting a deeper consideration of practical and system-specific aspects. These differences highlight the varied perspectives and interpretations brought by human expertise and AI analysis~\cite{different-perspective} in evaluating the effectiveness of MRs.

\section{Overall Evaluations of MRs generated by GPT-4}\label{sec:discussion-newMR}
Analyzing the evaluation results of MRs from human experts and GPT across three categories of systems—basic computational functions, complex systems without AI integration, and complex systems with AI integration—reveals some key similarities and differences in their assessment approaches and perspectives on the MRs.

\subsection{Overall summary of the MR evaluations across system categories}
Among the MRs generated by GPT-4 for the nine systems across three categories, there are some common aspects based on the evaluation results:
\begin{enumerate}
    \item Completeness and applicability of the MRs:\newline 
    The MRs across all system types are structured to include all necessary components, such as well-defined source test cases, input relations, and output relations. They effectively focus on and highlight key features and behaviors of the SUTs.

    \item Broad generalizability of the MRs:\newline 
    The MRs demonstrate wide generalizability across various systems. This trait is shown in both basic and complex systems, including those with AI components.

    \item Correctness of the MRs:\newline 
    The MRs generally provide accurate representations of intended system behaviors, a critical aspect recognized across various types of systems. However, there is a noted variation in the perceived level of correctness, particularly when it comes to complex and AI-embedded systems. To achieve higher levels of correctness, it is recognized that the MRs should have clearer-defined constraints to make the scenarios more accurate. These constraints are essential for ensuring that the MRs are not only correct, but also precise and directly applicable to the behaviors and specifications of the SUTs. This need for enhanced specificity and constraint is especially required for complex systems, where missing details could cause the MRs to violate the specifications of the SUTs.

    \item Novelty and clarity in the MRs:\newline 
    The novelty of MRs across different system categories is generally perceived as moderate rather than high. In basic math programs, the MRs are often based on fundamental arithmetic operations, which do not present a high level of novelty due to their foundational nature in mathematics. In complex systems, while some elements of innovation are acknowledged, the MRs are primarily viewed as extensions or adaptations of pre-existing concepts in the literature or human knowledge, rather than entirely novel ideas. This trend suggests that while the MRs incorporate new methods or perspectives in their approach to input and output relations, they primarily build upon established testing paradigms and techniques. The clarity of the MRs is generally high, but understanding them may require specific domain knowledge, especially in more complex applications.

    \item Computational feasibility of the MRs:\newline 
    The computational feasibility of implementing the MRs varies across system categories. In basic computational functions, the MRs are generally straightforward to implement and automate, owing to the simpler nature of the systems and the clear-cut operations involved. This allows for efficient generation and validation of test cases, making the MRs highly practical in these contexts. However, as the complexity of systems increases, the computational feasibility of the MRs has decreased. In these scenarios, the generation and automation of test cases can be more challenging. The complexity inherent in these systems often requires more sophisticated algorithms and computational resources. This is particularly true for systems with AI, where the unpredictability and intricacy of AI behaviors~\cite{AI-unpredictability} can complicate the automation process. While the potential for automating test case generation based on MRs is acknowledged, the evaluations indicate that automating the entire MT process for the MRs is hard and currently limited by the technology.
    
\end{enumerate}

The MRs generated by GPT-4 exhibit strengths in completeness, relevance, and broad generalizability, demonstrating the model's adeptness in generating useful MRs for various systems. However, the correctness of these MRs, particularly in complex environments, highlights areas for improvement through including more detailed and specific constraints. The novelty aspect of these MRs stands out as a blend of innovation and established methodologies, indicating GPT-4's capability to integrate new ideas within traditional frameworks, though suggesting room for further originality. The computational feasibility and system-specific design of these MRs vary with system complexity, reflecting both the capabilities and limitations of GPT-4 in generating contextually appropriate and effective testing strategies. In summary, while GPT-4's MRs are useful in system testing, particularly in standard applications, their effectiveness in complex or novel scenarios could be enhanced through more focused development and refinement.

\subsection{Observations of evaluators}
Human evaluators tend to focus more on the practical and intricate aspects of MRs. They are critical and detail-oriented, especially regarding system-specific correctness and practical feasibility.
GPT evaluation reflects a broader, possibly less detailed perspective. GPT tends to be more optimistic in its assessments, particularly regarding novelty, clarity, and computational feasibility.

\section{Discussion}\label{sec:discussion}

\begin{figure*}
\captionsetup{width=0.9\textwidth}
    \centering
    \includegraphics[width=0.9\textwidth]{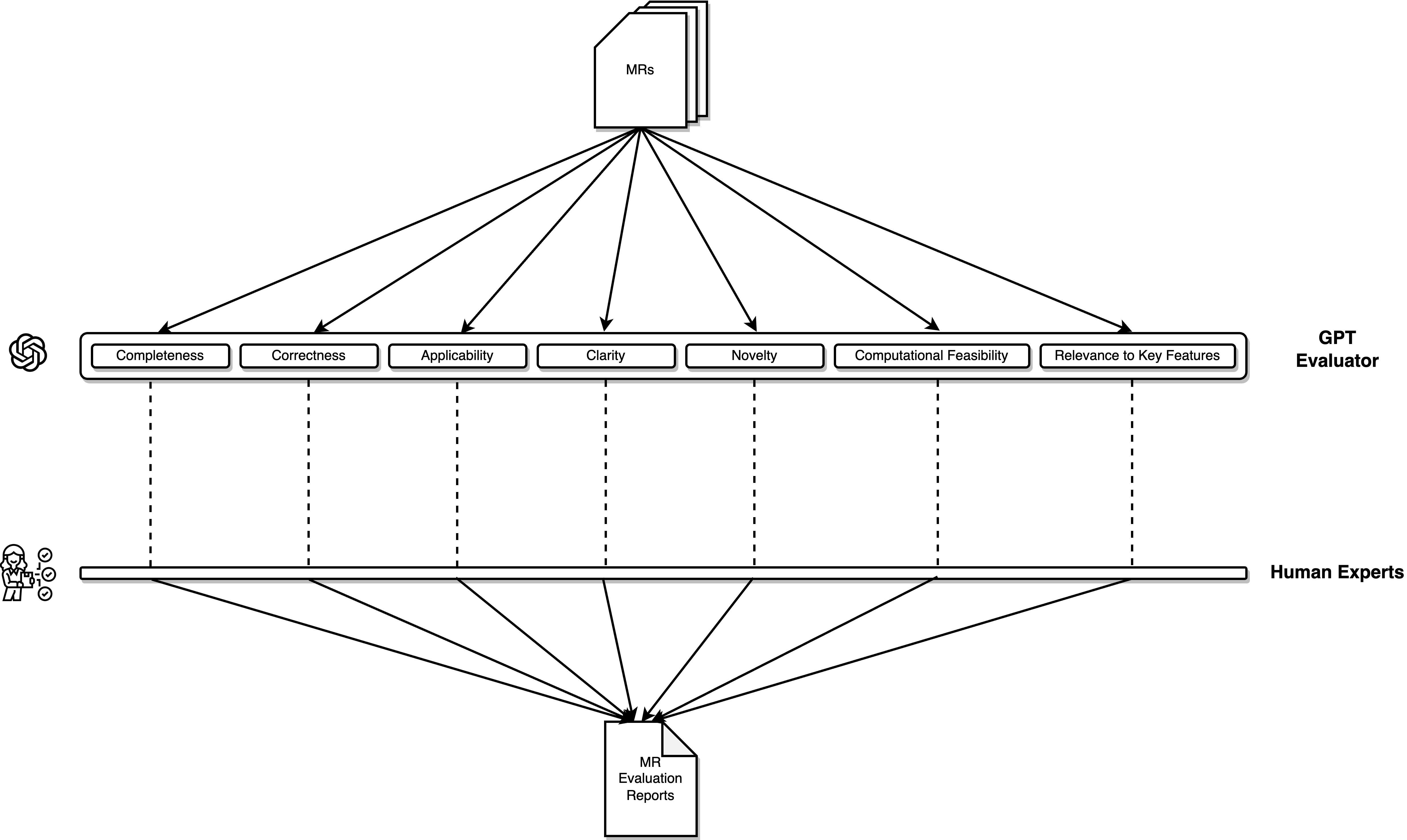}
    \caption{Human experts and ChatGPT work together to evaluate MRs}
    \label{fig:human-gpt-evaluate}
\end{figure*}

\subsection{Limitations}
This study's exploration into GPTs' ability to generate MRs for various SUTs are set against the fast development of GPT versions. While recognizing that each version of GPT might have a better performance, this rapid advancement does not diminish the relevance and significance of the current research findings. Rather, the current analysis of GPT-4's capabilities offers valuable benchmarks in the domain of MR generation. It is important to note, however, that although MRs generated by GPT could potentially be improved with more expertly crafted prompts, this approach requires more specialized knowledge and a higher level of proficiency in using GPT~\cite{human-knowledge-necessary}. Therefore, it falls outside the scope of this paper. This research aims to explore the performance of GPT in generating MRs with fewer prompts, providing a more focused evaluation of ChatGPT's capabilities.

Despite the rapidly evolving nature of specific GPT models like GPT-4, the methodologies and evaluation criteria developed in this study hold significant and lasting value. These methods offer a robust framework for assessing the generation of MRs in both current and future iterations of GPT models. As newer versions of GPT are released, the established criteria provide a consistent and reliable benchmark for comparison. This study, by evaluating the capabilities and limitations of GPT-4, offers critical findings that are immediately applicable and also lays the groundwork for assessing future advancements in GPT technology. Therefore, the findings from this research serve not only as a valuable snapshot of GPT-4's current performance but also as a foundational reference for future developments in GPT-based MR generation.

\subsection{Using GPT as a MR evaluator: strengths and future work}

\subsubsection{Strengths of GPT as an Evaluator}
Our study highlights several areas where GPT demonstrates notable strengths over human evaluators in MR assessment. Firstly, GPT exhibits an impressive capacity for rapidly evaluating a diverse range of MRs. This is particularly advantageous for large-scale evaluations, where the volume of MRs would be impractical for human evaluators to assess within a reasonable timeframe.

Additionally, GPT's evaluations are characterized by remarkable consistency, free from the biases and fatigue that may affect human evaluators. In terms of novelty, clarity, and computational feasibility, GPT consistently delivered objective and broad assessments. This capability of GPT could be especially beneficial in the initial screening processes of MRs, where quick and extensive evaluations are necessary.

\subsubsection{Limitations and Challenges of GPT as an Evaluator}
Our experiment results reveal limitations in GPT's evaluative precision, especially in its capacity to examine the finer details of MRs. A recurring observation was GPT's tendency to overlook the details of MRs when evaluating the correctness criterion, often resulting in higher scores. This tendency was obvious when evaluating MRs for large, complex systems, both with and without AI/ML components.

In contrast, human evaluators consistently demonstrated a more critical and detail-oriented approach, emphasizing the need for additional constraints to ensure MR correctness. This discrepancy indicates that while GPT can provide broad assessments, it sometimes struggles with the detailed and context-specific aspects that are crucial in evaluating complex systems, areas where human evaluators currently show greater adeptness.

However, it is important to note that such ability of human evaluators is largely dependent on their specific knowledge bases. In scenarios where human experts are not familiar with certain systems, GPT's evaluations, benefiting from its extensive training data encompassing a wide range of contexts, can be more advantageous. In such cases, GPT can provide insights or identify potential issues that might not be immediately apparent to human evaluators with limited exposure to the specific domain. This highlights the potential of GPT as a complementary tool in MR evaluation, particularly in unfamiliar or less-explored domains for human evaluators.

\subsubsection{Complementary Roles of GPT and Human Evaluators}
Given these strengths and limitations, we propose a synergistic model combining the capabilities of GPT with human expertise. Although GPT is capable of effectively handling the initial, high-level assessment of MRs, its inclination to disregard specific details requires a further, more thorough examination by human evaluators. This is especially critical for complex systems, where the depth and specificity of human judgement are essential to identify and rectify potential oversights by GPT.

Therefore, as Figure~\ref{fig:human-gpt-evaluate} shows, GPT could serve as a first-line evaluator, rapidly assessing and filtering MRs based on broad criteria such as novelty and computational feasibility. Subsequently, human evaluators, with their attention to detail and critical perspective, could then thoroughly assess these MRs, ensuring that additional constraints are identified and applied to guarantee certain aspects, such as correctness, particularly in large and complex test cases. This hybrid approach not only maximises efficiency but also leverages the unique strengths of both AI and human intelligence, leading to a more robust and comprehensive evaluation process.

\subsubsection{Areas for improvement in GPT evaluation}

The observed tendency of GPT to overlook MR details and overestimate correctness highlights a need for improvement in both the evaluation criteria and GPT configurations. To better align with human evaluators, GPT could benefit from integrating more sophisticated, context-aware evaluation metrics or training on more diverse and complex datasets. Further, integrating algorithms or configurations that mimic human-like critical thinking and attention to detail could improve its evaluative accuracy. 

\section{Conclusion}\label{sec:conclusion}
This study conducted a comparative analysis of the MRs generated by GPT-3.5 and GPT-4, using established and then updated evaluation criteria. The initial comparison explicitly demonstrated GPT-4's superiority in generating higher-quality MRs compared to GPT-3.5. Further, applying the refined criteria to a broader range of nine SUTs, including both simple and AI/ML complex problems, provided a more comprehensive and thorough analysis of the quality of the MRs generated by GPT-4. A novel aspect of this study was the use of both a customized GPT evaluator and the human evaluators, offering a comparative insight into the capabilities of automated versus human assessment of the MRs.

The findings demonstrate GPT-4's advanced capabilities in software testing and MR generation across diverse applications. It highlights the evolving capabilities of AI in software testing, especially in generating and assessing MRs, while also emphasizing the indispensable role of human expertise in critical and detail-oriented evaluation processes.

\section*{Acknowledgments}
This work was supported by the 2025 Key Technological Innovation Program of Ningbo City under Grant No. 2022Z080, and in part by the Australian Research Council Project under Grant DP210102447.

\bibliographystyle{IEEEtran}
\bibliography{main}

\end{document}